\documentclass[12pt]{article}
\usepackage{amsfonts,amsmath}
\usepackage[mathscr]{eucal}

\def\theequation{\arabic{section}.\arabic{equation}}
\newcommand{\be}{\begin{equation}}\newcommand{\ee}{\end{equation}}
\newcommand{\bea}{\begin{eqnarray}}
\newcommand{\eea}{\end{eqnarray}}

\newcommand{\lb}{\label}
\newcommand{\p}[1]{(\ref{#1})}

\begin{document}

\begin{titlepage}

\begin{flushright}
ITP-UH-07/12
\end{flushright}

\vspace*{0.2cm}

\begin{center}

{\LARGE\bf Nahm equations in supersymmetric mechanics}

\vspace{2.0cm}

\renewcommand{\thefootnote}{$\star$}

{\large\bf Sergey~Fedoruk\,\footnote{\,\,On leave of absence
from V.N.\,Karazin Kharkov National University, Ukraine}
,\quad
Evgeny~Ivanov}

\vspace{0.5cm}

{\it Bogoliubov Laboratory of Theoretical Physics, JINR,}\\
{\it 141980 Dubna, Moscow region, Russia} \\

\vspace{0.1cm}

{\tt fedoruk,eivanov@theor.jinr.ru}

\vspace{1.0cm}

{\large\bf Olaf~Lechtenfeld}

\vspace{0.5cm}

{\it Institut f\"ur Theoretische Physik,
Leibniz Universit\"at Hannover,}\\
{\it Appelstra{\ss}e 2, D-30167 Hannover, Germany} \\

\vspace{0.1cm}

{\tt lechtenf@itp.uni-hannover.de}

\end{center}

\vspace{2.0cm}

\begin{abstract}
\noindent
We elaborate on a novel model of ${\cal N}{=}\,4$
supersymmetric mechanics with extra spin variables.
A dynamical linear ({\bf 1,4,3}) multiplet is coupled to a
``semi-dynamical'' linear ({\bf 3,4,1}) multiplet representing
spin degrees of freedom in a Wess-Zumino action. The unique coupling
of these two multiplets relates the dynamical bosonic variable
to an arbitrary harmonic function of the SU(2) triplet of spin variables.
As we prove at the classical and quantum level, ${\cal N}{=}\,4$
supersymmetry is equivalent to the Nahm equations for the spin variables,
with the dynamical boson as evolution parameter. We treat in detail the
one- and two-monopole as well as some special multi-monopole configurations.
While one monopole exhibits superconformal OSp(4$|$2) symmetry
and was worked out previously, only ${\cal N}{=}\,4, d\,{=}1\,$ Poincar\'e
supersymmetry survives for multi-monopole configurations.
\end{abstract}

\vfill

{\small

}

\newpage

\end{titlepage}

\setcounter{footnote}{0}

\setcounter{equation}0
\section{Introduction}

Recently, a new type of models for ${\cal N}{=}\,4$ supersymmetric mechanics was
discovered and studied \cite{FIL1,FIL2,FIL3,FIL-r}. They are distinguished from earlier such models 
by the coupling of two irreducible ${\cal N}{=}\,4$ multiplets, one dynamical and one ``semi-dynamical''.
The former produces normal kinetic terms for all components, while the kinetic terms of the latter
are one order lower in time derivatives: the bosonic variables arise in a $d\,{=}\,1$ Wess-Zumino
(WZ) action, and the fermionic variables appear only algebraically in the action, thus are auxiliary.
After quantization, the semi-dynamical bosonic variables play the role of spin degrees of freedom
parametrizing a fuzzy manifold.~\footnote{
In the simplest case of ${\rm SU(2)}$ doublets one gets the standard fuzzy sphere~\cite{Mad}.}
For this reason, we also call the semi-dynamical multiplet the ``spin multiplet''. 
A slightly different treatment of the semi-dynamical spin variables was employed in~\cite{BK1,KL,KL3}.

The first examples of these compound ${\cal N}{=}\,4$ supersymmetric mechanics models
were constructed in \cite{FIL2,FIL3} as a one-particle limit of a new type of
${\cal N}{=}\,4$ super Calogero models \cite{FIL1}. They describe an off-shell coupling of a
dynamical ({\bf 1,4,3}) multiplet to a gauged ({\bf 4,4,0}) spin multiplet.
They inherit the superconformal $D(2,1;\alpha)$ invariance of the parent super Calogero models~\cite{KL1}
and realize a novel mechanism of generating a conformal potential $\sim x^{-2}$ for the dynamical
bosonic variable, with a quantized strength. Soon after, the construction was generalized by replacing 
the dynamical ({\bf 1,4,3}) multiplet with a ({\bf 4,4,0}) or a ({\bf 3,4,1}) one, but still keeping 
the ({\bf 4,4,0}) spin multiplet \cite{BKS,ISKon,KL2,IKon}.
The larger number of dynamical bosons allowed for Lorentz-force-type couplings to non-abelian self-dual
background gauge fields in a manifestly ${\cal N}{=}\,4$ supersymmetric fashion \cite{ISKon,IKon}.
Here, the presence of the spin variables is essential for going beyond abelian backgrounds.
It has been conjectured in~\cite{FIL2,FIL3} that these compound supersymmetric mechanics models
may be also relevant to the description of ${\cal N}{=}\,4$ supersymmetric black holes.

In the present paper, we entertain a different generalization of the ({\bf 1,4,3})--({\bf 4,4,0}) model,
by replacing the ({\bf 4,4,0}) spin multiplet with a linear ({\bf 3,4,1}) multiplet. For the dynamical
multiplet, we remain with the ({\bf 1,4,3}) one, postponing other choices to future study.
We employ the ${\cal N}{=}\,4, d\,{=}\,1$ harmonic superspace approach~\cite{GIOS,IL} for the off-shell
description of the ${\cal N}{=}\,4$ multiplets. To keep the treatment as general as possible, we
require only ${\cal N}{=}\,4, d\,{=}\,1$ Poincar\'e invariance, which includes ${\cal N}{=}\,4$ superconformal
systems as a subclass.

What is the effect of changing the spin multiplet?
In the ({\bf 4,4,0}) case~\cite{FIL2,FIL3}, the bosonic
variables form ${\rm SU(2)}$ doublets and after quantization span an oscillator-type Heisenberg algebra.
The fuzzy sphere arises from applying a quantum version of the $S^{\,3} \rightarrow S^{\,2}$ Hopf fibration.
In the ({\bf 3,4,1}) case considered here, in contrast, the elementary bosonic variables $v_a$, $a=1,2,3\,$,
form an SU(2) triplet. The WZ action for the ({\bf 3,4,1}) multiplet produces scalar $\mathscr{U}(v)$ and
vector $\mathscr{A}_{a}(v)$ potentials in the component action, which are related by the four-dimensional
self-duality equation ${\rm rot}\,\vec{\mathscr{A}}={\rm grad}\,\mathscr{U}$.
The scalar potential $\mathscr{U}$ must be a harmonic function in $\mathbb{R}^3\ni\{v_a\}$ and
is related to the dynamical bosonic variable $x$ of the ({\bf 1,4,3}) multiplet by the constraint $\mathscr{U}=x$
from the superfield coupling of the two multiplets.
As a result, only two bosonic degrees of freedom remain independent in the ({\bf 3,4,1}) multiplet.
Being semi-dynamical (i.e.~of first order in time derivatives), these are the genuine spin variables.

This separation of degrees of freedom carried by the three-vector $v_a$ has a remarkable consequence.
The set of Hamiltonian constraints implies a variant of the celebrated Nahm equations for the vector $v_a$,
with the dynamical combination of the latter's components as the corresponding evolution parameter.
These Nahm equations play a very fundamental role: they represent the necessary and sufficient conditions
for ${\cal N}{=}\,4, d\,{=}\,1$ Poincar\'e supersymmetry in our model.
This phenomenon persists in the quantum theory: the ${\cal N}{=}\,4$ supercharges and quantum Hamiltonian
constitute the Poincar\'e superalgebra if and only if the quantum operators $\hat v_a$ satisfy the operator
version of the Nahm equations.

The two spin degrees of freedom encoded in the vector $v_a$ are described covariantly by a constrained
three-vector $\ell_a$, $a=1,2,3$. The constraint $x=\mathscr{U}$ and the Dirac brackets of $\ell_a$
then determine the geometry of the spin space. In the one- and special multi-monopole configurations
we shall consider, the spin variables describe a fuzzy two-sphere.
After canonical quantization \`a la Dirac, they {\it directly\/} yield an SU(2) algebra.~\footnote{
In contrast to the $({\bf 4,4,0})$ spin multiplet case,
where the spin variables form a Heisenberg algebra upon quantization, while the fuzzy sphere
and SU(2) group are recovered through a quantum Hopf fibration~\cite{FIL2}.}
For the one-monopole situation, these features could actually be expected from the results of~\cite{FIL2}
by way of a special (non-Wess-Zumino) gauge choice.
In the two-monopole case, only a U(1) symmetry survives, and we propose to use as a canonically conjugated
pair of spin variables the polar angle $\varphi=\arctan (\ell_2/\ell_1)$ and the U(1) generator~$\ell_3$.
This choice allows us to take advantage of the Moyal-bracket formalism for the Weyl-ordered quantum variables.

The WZ action used to describe the ({\bf 3,4,1}) spin multiplet
has a consistent off-shell superfield realization only within ${\cal N}{=}\,4$ harmonic superspace.
For this reason, in Section~2 we start with the harmonic superfield action of the coupled
({\bf 1,4,3})--({\bf 3,4,1}) system and then derive the corresponding component action.
In Section~3 we analyze the bosonic limit of this system. We present the Hamiltonian formulation
for a general scalar potential and show how the classical SU(2) Nahm equations appear in this framework.
In the one- and two-monopole configurations we discuss in detail the definition of the spin variables
and perform the quantization. The full supersymmetric systems are considered in Section~4.
We explain the relationship between ${\cal N}{=}\,4$ supersymmetry and the Nahm equations at the
classical and quantum level. Section~5 provides a short summary of our results and discusses peculiarities
of the multi-monopole situation.

\setcounter{equation}0
\section{Superfield content and action}

We shall deal with the off-shell superfield action
\begin{equation}\label{A}
S = S_{{\mathscr{X}}} +S_{int} +S_{WZ}= \displaystyle{\int} \mu_H \,
\mathscr{L}({\mathscr{X}}) + {\textstyle\frac{i}{2}} \,b
\displaystyle{\int} \mu^{(-2)}_A \, \mathcal{V}\, (L^{++}+c^{++})
-{\textstyle\frac{i}{2}} \,\gamma\displaystyle{\int} \mu^{(-2)}_A \,
\mathscr{L}^{(+2)} (L^{++},u)\,.
\end{equation}
Here, $c^{++} =c^{ik}u^+_iu^+_k\,$. The renormalization constants $\gamma$ and $b$, as well as the constant
triplet $c^{ik}$, are parameters of the model. Below they will be appropriately fixed, either
by redefining component fields or by using some (broken) symmetries. For instance, the norm of the vector $c^{ik}$
can be fixed at any non-zero value by properly rescaling $L^{++}$ and the parameter $b$.

The ${\cal N}=4$ superfields ${\mathscr{X}}$ and $L^{++}$
accommodate the off-shell ({\bf 1,4,3}) and ({\bf 3,4,1}) multiplets,
their precise definition is given below. The analytic
superfield $\mathcal{V} = \mathcal{V}(\zeta, u)$ is the prepotential
for the ({\bf 1,4,3}) multiplet related to the superfield
${\mathscr{X}}(t,\theta_i,\bar\theta^i)$ by the harmonic integral
transform~\cite{DI1}
\begin{equation} \label{X-V}
{\mathscr{X}}(t,\theta_i,\bar\theta^i)=\int du
\,\mathcal{V}(t_A,\theta^+,\bar\theta^+,u)\Big|_{\theta^\pm=\theta^i u^\pm_i,\,\,\,
\bar\theta^\pm=\bar\theta^i u^\pm_i} \,.
\end{equation}
The definition~(\ref{X-V}) is invariant under the gauge transformation
\begin{equation}\label{Nu-gauge-tr}
\delta \mathcal{V} = D^{++}\Lambda^{--}\,, \qquad \Lambda^{--}=\Lambda^{--}(\zeta,u)\,.
\end{equation}

The term proportional to $c^{++}$ in~(\ref{A}), i.e. $i
\displaystyle{\int} \mu^{(-2)}_A \, \mathcal{V}\, c^{++}$, is the FI term for the
{\bf (1,4,3)} multiplet.

Next two subsections contain a brief characterization of the
linear ({\bf 1,4,3}) and ({\bf 3,4,1}) multiplets.

\subsection{The ({\bf 1,4,3}) multiplet}

The analytic prepotential formulation of the ({\bf 1,4,3}) multiplet as given in \p{X-V}, \p{Nu-gauge-tr}
was proposed in \cite{DI1} in the framework of the general ${\cal N}{=}4, d\,{=}1$ superfield gauging procedure \cite{DI}.
This formulation was recently employed in \cite{FIL2,FIL3} in the context close to the subject
of the present paper.

In the ordinary ${\cal N}{=}4$ superspace parametrized by the coordinates $\theta_i$,
$\bar\theta^i$ and $t$ the ({\bf 1,4,3}) multiplet is described by the superfield
$\mathscr{X}(t,\theta_i,\bar\theta^i)$ subjected to the constraints~\cite{IKL}
\begin{equation} \label{cons-X-c}
D^iD_i \,\mathscr{X}=0\,, \qquad \bar D_i\bar D^i \,\mathscr{X}=0\,, \qquad [D^i,\bar
D_i]\, \mathscr{X}=0\,,
\end{equation}
where $D^i=\partial/\partial\theta_i-i\bar\theta^i\partial_t$, $\bar D_i=\partial/\partial\bar\theta^i-i\theta_i\partial_t$
are spinor covariant derivatives\footnote{
Our ${\cal N}{=}4$ superspace conventions are the same as in refs. \cite{ISKon,IKon} and in our recent review \cite{FIL-r}.
They differ from those used, e.g., in \cite{IL,DI,DI1} by the sign of the evolution parameter $t$.
The advantage of this choice is that it directly yields the correct sign of the fermionic kinetic term and, consequently,
the sign `plus' in front of the right-hand side of the quantum supersymmetry algebra anticommutator, $\{Q,Q^\dagger\}=2H\,$.
Quantization of ${\cal N}{=}4$ supersymmetric mechanics with the conventions of \cite{IL,DI,DI1} can be found in \cite{FIL3}.
}. The $\theta$ expansion of this superfield is as follows
\begin{equation} \label{sing-X0-WZ}
\mathscr{X}(t,\theta_i,\bar\theta^i)= x + \theta_i\chi^i + \bar\chi_i\bar\theta^i +
\theta^i\bar\theta^k K_{ik}-{\textstyle\frac{i}{2}}(\theta)^2\dot{\chi}_i\bar\theta^i
-{\textstyle\frac{i}{2}}(\bar\theta)^2\theta_i\dot{\bar\chi}{}^i +
{\textstyle\frac{1}{4}}(\theta)^2(\bar\theta)^2 \ddot{x}\,,
\end{equation}
with $(\theta)^2\equiv \theta_i\theta^i$, $(\bar\theta)^2\equiv
\bar\theta^i\bar\theta_i\,$. The first term in the action (\ref{A}),
$
S_{\mathscr{X}} =
\displaystyle{\int} dt \,d^4\theta \, \mathscr{L}({\mathscr{X}})\,,
$
where
$d^4\theta=\frac{1}{4}\frac{\partial}{\partial\bar\theta_i}\frac{\partial}{\partial\bar\theta^i}
\frac{\partial}{\partial\theta^i}\frac{\partial}{\partial\theta_i}$, has the following
component form
\begin{equation}\label{A-1-0-com}
S_{\mathscr{X}} = \int dt \, \Big[ \mathscr{L}^{\prime}\ddot{x} -
i\mathscr{L}^{\prime\prime} \left(\dot{\bar\chi}{}^k \chi_k - {\bar\chi}^k
\dot\chi_k\right)
  + {\textstyle\frac12}\,\mathscr{L}^{\prime\prime}K^{ik} K_{ik}
- \mathscr{L}^{\prime\prime\prime}K^{ik}\chi_i \bar\chi_k +
{\textstyle\frac{1}{4}}\,\mathscr{L}^{({IV})}\chi_i\chi^i \bar\chi^k
\bar\chi_k\Big]\,.
\end{equation}
Here, $\mathscr{L}^{\prime}$, $\mathscr{L}^{\prime\prime}$,
$\mathscr{L}^{\prime\prime\prime}$, $\mathscr{L}^{({IV})}$ are functions of $x\,$,
and primes mean differentiation with respect to $x$: $\mathscr{L}^{\prime}=\mathscr{L}^{\prime}(x)$, etc.

It is easy to see that the prepotential representation~(\ref{X-V}) solves the
constraints~(\ref{cons-X-c}). Expressing $D_i=D^- u^+_i -D^+ u^-_i$ and $\bar D_i=\bar D^-
u^+_i -\bar D^+ u^-_i$ and using the only non-vanishing anticommutation relations
\be
\{ D^+,\bar D^-\}=-\{
D^-,\bar D^+\}=-2i\partial_t\,,
\ee
we find that the constraints~(\ref{cons-X-c}) are satisfied as a direct consequence of the analyticity
conditions for $\mathcal{V}$
\begin{equation} \label{constr-V0}
{D}^{+} \,\mathcal{V}=\bar{D}^{+}\, \mathcal{V}=0\,.
\end{equation}

To be convinced that $\mathcal{V}$ indeed describes the ({\bf 1,4,3}) multiplet we need
to exploit the gauge freedom \p{Nu-gauge-tr}. It can be used to remove an infinite set of
gauge degrees of freedom from $\mathcal{V}$ and to bring it into the Wess-Zumino gauge form
\begin{equation} \label{V0-WZ}
\mathcal{V} (t_A, \theta^+, \bar\theta^+, u^\pm) =x(t_A)- 2\,\theta^+ \chi^{i}(t_A)u^-_i -
2\,\bar\theta^+ \bar\chi^{i}(t_A)u^-_i + 3\,\theta^+ \bar\theta^+ K^{ik}(t_A)u^-_i u^-_k
\,,
\end{equation}
where $t_A=t+i(\theta^+\bar\theta^-+\theta^-\bar\theta^+)$. The fields $x(t)$, $\chi^{i}(t)$, $\bar\chi^{i}(t)$, $K^{ik}(t)$ are the same as in
\p{sing-X0-WZ}. To preserve this gauge, the standard ${\cal N}=4$ supersymmetry
transformations of $\mathcal{V}$ should be accompanied by a proper field-dependent gauge transformation
with a composite $\Lambda^{--}$.

\subsection{The linear (3,4,1) multiplet and its WZ action}

The linear ${\cal N}{=}4$ multiplet is accommodated by the even analytic gauge superfield
$L^{++}(\zeta,u)$ subjected to the additional harmonic constraint \cite{IL}
\begin{equation}\label{con-lm}
{D}^{++} \,L^{++}=0\,.
\end{equation}

The constraints \p{con-lm} can be directly solved.
The off-shell component content of the linear multiplet is comprised by the fields $v_{ij} = v_{ji}$, $B$,
$\psi_i$ and $\bar\psi_i$. They enter the $\theta$ -expansion of the
superfield $L^{++}$ subjected to \p{con-lm} as \cite{IL}
\begin{equation}\label{sol-LV-com}
L^{++} = v^{++} +\theta^+ \psi^+ +\bar\theta^+ \bar\psi^+ -  2i \,\theta^+ \bar\theta^+
\left(\dot v^{+-} +B \right)\,,
\end{equation}
where $v^{++}=v^{ij}u^+_iu^+_j\,, \;v^{+-}=v^{ij}u^+_iu^-_j$, $\psi^+=\psi^i u^+_i$ and $\bar\psi^+=\bar\psi^i u^+_i$.

When taken separately, the last (WZ) term in the action (\ref{A}), $S_{WZ}$, provides an example
of supersymmetric Chern-Simons mechanics \cite{FJ,FRS,HTown}. In components, it
has the following form
\begin{eqnarray}\label{A-WZ-com-b-lin}
S_{WZ} = -\gamma \int dt \,du \, \frac{\partial \mathscr{L}^{(+2)} (v^{++},u)}{\partial
v^{++}} \,\Big(\dot v^{+-}+ B\Big)- {\textstyle\frac{i}{2}}\,\gamma \int dt \,du \, \frac{\partial^2
\mathscr{L}^{(+2)} (v^{++},u)}{\partial (v^{++})^2}\,\, \bar\psi^+\psi^+\,.\nonumber
\end{eqnarray}
It can be rewritten as
\begin{equation}\label{WZ-com-b-lin}
S_{WZ} = -\gamma \int dt \, \Big({\textstyle\frac{1}{2}}\,\mathscr{A}_{ik}\dot v^{ik} +
{\textstyle\frac{i}{2}}\, \,\mathscr{R}_{ik}\bar\psi^{(i}\psi^{k)} + \mathscr{U} B \Big),
\end{equation}
where
\begin{equation}\label{pot-lin}
\mathscr{A}_{ik}= 2\int du \,u^+_{(i}u^-_{k)}\, \frac{\partial \mathscr{L}^{++} }{\partial
v^{++}}\,,\qquad \mathscr{R}_{ik}= \int du \,u^+_{i}u^+_{k}\, \frac{\partial^2
\mathscr{L}^{++} }{\partial (v^{++})^2}\,.\qquad \mathscr{U}=\int du \, \frac{\partial
\mathscr{L}^{++} }{\partial v^{++}}\,.
\end{equation}
{}From the definition of these potentials follow the relations between them:
\begin{equation}\label{const-pot-lin1}
\triangle_{\mathbb{R}^3}\mathscr{U}=0\,,\qquad
\triangle_{\mathbb{R}^3}\mathscr{A}_{ik}=0\,,\qquad
\partial^{ik}\mathscr{A}_{ik}=0\,,
\end{equation}
\begin{equation}\label{const-pot-lin2}
\partial_{ij}\mathscr{A}_{kl}-\partial_{kl}\mathscr{A}_{ij}=
\left(\epsilon_{ik}\partial_{jl}+\epsilon_{jl}\partial_{ik}\right)\mathscr{U}\,,
\end{equation}
\begin{equation}\label{const-pot-lin3}
\mathscr{R}_{ik}=
\partial_{ik}\mathscr{U}\,.
\end{equation}
Here, $\partial_{ik}=\partial/\partial v^{ik}$ and $
\triangle_{\mathbb{R}^3}=\partial^{ik}\partial_{ik} $ is Laplace operator on
$\mathbb{R}^3$.

Eqs. (\ref{const-pot-lin1}), (\ref{const-pot-lin2}) are recognized as the equations
defining the monopole (static) solution for a self-dual Maxwell or gravitation fields in $\mathbb{R}^4$
(see, for example, \cite{EGHan} and refs. therein).
Namely, the non-trivial physics arises in the presence of singularities in $\mathscr{U}$.
These singularities lead to a non-trivial definition of the background vector gauge potential $\mathscr{A}_{ij}$
which necessarily involves the Dirac strings.
This property requires the use of multiple covering (or the fibre bundle formalism) of the $v$--space
for the correct definition of the potential \cite{EGHan}.
Although we will  sometimes give the expression for $\mathscr{A}_{ij}$,
analysis of our system does not require knowledge of the exact expression for this vector potential.
The Hamiltonian analysis of our system will deal solely with the field strengths.

\subsection{The $\mathcal{V}$ - $L^{++}$ interaction}

The second term in (\ref{A}), $S_{int}$, describes an interaction of the ({\bf 1,4,3}) and ({\bf 3,4,1})
multiplets. Its form is uniquely determined by requiring it to be invariant under the gauge
transformations \p{Nu-gauge-tr}. It has the following simple component representation
\begin{eqnarray}\label{4N-WZ}
S_{int} &=& {\textstyle\frac{i}{2}} \,b\int \mu^{(-2)}_A \, \mathcal{V}\, (L^{++}+c^{++})\\
&=&
ib\int dt \,\Big[ -ix \,B
+{\textstyle\frac{1}{2}} \left( \bar \chi^k \psi_k- \bar \psi^k \chi_k\right) +
{\textstyle\frac{1}{2}} \,K^{ij} (v_{ij}+c_{ij}) \Big].\nonumber
\end{eqnarray}

\subsection{The total component action}

After summing up the component actions \p{A-1-0-com}, \p{WZ-com-b-lin} and \p{4N-WZ}, the
total action (\ref{A}) in terms of the component fields reads
\begin{eqnarray}
S &=& \int dt \, \Big[ \mathscr{L}^{\prime}\ddot{x}- i\mathscr{L}^{\prime\prime}
\left(\dot{\bar\chi}{}^k \chi_k - {\bar\chi}^k \dot{\chi}_k\right) +
{\textstyle\frac12}\,\mathscr{L}^{\prime\prime}K^{ik} K_{ik} -
\mathscr{L}^{\prime\prime\prime}K^{ik}\chi_i \bar\chi_k +
{\textstyle\frac{1}{4}}\,\mathscr{L}^{({IV})}\chi_i\chi^i
\bar\chi^k \bar\chi_k \nonumber\\
&& \qquad +\, b\,x B +{\textstyle\frac{i}{2}} \,b\left( \bar \chi^k \psi_k- \bar \psi^k \chi_k\right)
+{\textstyle\frac{i}{2}} \,b\,K^{ij} (v_{ij}+c_{ij})\label{A-com-lin}\\
&& \qquad -\,{\textstyle\frac12}\,\gamma \mathscr{A}_{ik}\dot v^{ik}-
{\textstyle\frac{i}{2}}\,\gamma \,\mathscr{R}_{ik}\bar\psi^{(i}\psi^{k)} - \gamma
\,\mathscr{U}\,B \Big].\nonumber
\end{eqnarray}

Next we should use the algebraic equations of motion for the auxiliary fields $K_{ik}$, $\psi_{k}$ and $\bar\psi_{k}$
$$
K_{ik}=(\mathscr{L}^{\prime\prime})^{-1}\left[ \mathscr{L}^{\prime\prime\prime}\chi_{(i} \bar\chi_{k)}
-{\textstyle\frac{i}{2}} \,b\, (v_{ik}+c_{ik})\right]\,,\qquad
\psi_{i}=-b\gamma^{-1}(\mathscr{R}^{-1})_{ik}\chi^{k}\,,
\quad \bar\psi_{i}=-b\gamma^{-1}(\mathscr{R}^{-1})_{ik}\bar\chi^{k}\,,
$$
where $(\mathscr{R}^{-1})^{ik}=2\mathscr{R}^{ik}/(\mathscr{R}^{lm}\mathscr{R}_{lm})$ is inverse of $\mathscr{R}_{ik}$
defined in \p{const-pot-lin3}.
Integrating out these auxiliary fields from \p{A-com-lin}, we obtain the action in terms of physical fields only:
\begin{eqnarray}
S &=& \int dt \, \Big[ \mathscr{L}^{\prime}\ddot{x}
+ {\textstyle\frac{1}{8}}\,b^2(\mathscr{L}^{\prime\prime})^{-1}(v^{ij}+c^{ij}) (v_{ij}+c_{ij})
-{\textstyle\frac12}\,\gamma \mathscr{A}_{ik}\dot v^{ik}
- (\gamma \,\mathscr{U}-b\,x)B \nonumber\\
&&  \qquad -\,
i\,\mathscr{L}^{\prime\prime}
\left(\dot{\bar\chi}{}^k \chi_k - {\bar\chi}^k \dot{\chi}_k\right)
- {\textstyle\frac{i}{2}} \,b \Big( (\mathscr{L}^{\prime\prime})^{-1}\mathscr{L}^{\prime\prime\prime}
(v^{ik}+c^{ik}) + b\gamma^{-1}(\mathscr{R}^{-1})^{ik} \Big) \chi_i \bar\chi_k
\qquad \label{A-com-lin-1}\\
&& \qquad +\, {\textstyle\frac{1}{4}}\Big(\mathscr{L}^{(IV)} -
{\textstyle\frac{3}{2}}\,(\mathscr{L}^{\prime\prime})^{-1}(\mathscr{L}^{\prime\prime\prime})^{2} \Big)
\chi_i\chi^i \bar\chi^k \bar\chi_k
\Big].\nonumber
\end{eqnarray}

The renormalization constants $b$ and $\gamma$ mark the contributions of the superfield interaction and WZ
terms to the physical component action. They can be converted into some non-zero numbers by a proper rescaling
of the variables $v_{ik}$, $B$ and the potential $\mathscr{U}$. Hereafter, we set $b=1$, $\gamma=1$.

\setcounter{equation}0
\section{Bosonic limit}

\subsection{Lagrangian, Hamiltonian and constraints}

It is instructive to look first at the bosonic limit of the action \p{A-com-lin-1}. It reads:
\begin{equation}\label{A-bcom-lin1}
S^{bose} = \int dt \, \Big[ -\mathscr{L}^{\prime\prime}\dot{x}\dot{x} +
{\textstyle\frac{1}{8}}\,(\mathscr{L}^{\prime\prime})^{-1}(v^{ij}+c^{ij}) (v_{ij}+c_{ij})
-{\textstyle\frac12}\, \mathscr{A}_{ik}\dot v^{ik} + B\,(x - \mathscr{U})
\Big].
\end{equation}

We see that the effect of adding the superfield coupling $S_{int}$ between the
({\bf 1,4,3}) and ({\bf 3,4,1}) multiplets is two-fold: first, there appears an oscillator-type
potential term for the bosonic fields $v^{ik}$ of the ({\bf 3,4,1}) multiplet (with the
additional dependence on $x$) and, second, the auxiliary field $B$ appears as a Lagrangian
multiplier for the constraint
\begin{equation} \label{Constr}
x -  \mathscr{U}(v) = 0\,.
\end{equation}
Here, the first term comes from $S_{int}$, while the second one from the ({\bf 3,4,1}) WZ term $S_{WZ}\,$.

If we would leave, for the ({\bf 3,4,1}) multiplet, the WZ action $S_{WZ}$ alone we would
obtain the meaningless condition $\mathscr{U} = 0$. On the contrary, the constraint
\p{Constr} is quite reasonable, expressing bosonic field of one ${\cal N}=4$ multiplet (viz. ({\bf{1,4,3}}))
through a function of bosonic fields of another ({\bf{3,4,1}}) multiplet. Also, the potential term for
$v^{ik}$ arises as a result of elimination of the auxiliary fields $K^{ik}$ of the
({\bf 1,4,3}) multiplet. Thus we observe a new mechanism of producing the potential
terms in such a coupled system. Even more interesting, after substituting \p{Constr} into
the kinetic term of $x$, we finally obtain a non-trivial target metric
\begin{equation} \label{indMetr}
\sim \partial_{ik}\mathscr{U}\partial_{jl}\mathscr{U}\dot{v}{}^{ik}\dot{v}^{jl},
\end{equation}
while originally there was no any kinetic term for $v^{ik}$, only the WZ term. This
situation should be contrasted with the ({\bf 3,4,1}) supersymmetric mechanics models considered in
\cite{IKLecht,IL}, in which the invariant superfield actions from the very beginning involve both the
kinetic and WZ terms for the ({\bf 3,4,1}) multiplet. In this kind of ${\cal N}{=}4$ mechanics models
the kinetic term of $v^{ik}$ appears in parallel
with a term bilinear in $B$, and elimination of $B$ by its algebraic equation of motion generates a
potential $\sim (\mathscr{U})^2\,$. No any additional contribution to the target $v^{ik}$
metrics comes from the WZ terms in this case. The target space metric for one linear
({\bf 3,4,1}) multiplet is always conformally-flat, while the induced metric \p{indMetr} for
generic $\mathscr{U}$ does not feature this property.

However, the metric \p{indMetr} is clearly degenerated. One can pass to the new
parametrization of the target space, treating $\mathscr{U}$ as one of the new coordinates.
Then two remaining coordinates will not appear in the metric part at all and will
contribute only the WZ coupling. Thus there remains only one genuine dynamical coordinate
and two independent spin variables.
Taking this into account, for quantization it proves more
convenient not to explicitly solve the constraint \p{Constr} at all, viewing $x$ as an independent
phase variable. Eq. \p{Constr} will be treated as a second-class hamiltonian constraint, on
equal footing with some other second-class constraints associated with the action
(\ref{A-com-lin-1}).

To simplify things, in what follows we focus on the option with
$\mathscr{L}\vert=-\frac{1}{2}x^2$ (here, $\vert$ denotes restriction to the $\theta$-independent parts).
In terms of superfields, it corresponds to the particular choice $\mathscr{L}({\mathscr{X}}) = -\frac{1}{2}{\mathscr{X}}^2$ in \p{A}.
The action (\ref{A-bcom-lin1}) takes the form
\begin{equation}\label{ac-bcom-lin1}
S^{bose} = \int dt \, \Big[ \dot{x}\dot{x} - {\textstyle\frac{1}{8}}\,(v^{ij}+c^{ij})
(v_{ij}+c_{ij}) -{\textstyle\frac{1}{2}}\, \mathscr{A}_{ik}\dot v^{ik} + B\left( x- \mathscr{U}\right) \Big]\,.
\end{equation}

Let us introduce the three--vector notation, passing from the spinor triplet indices $(ik)$
to the vector ones $a = 1,2,3$:
\be
v^{ik}=i\sigma_{a}^{ik}v_a\,,\qquad
v_a={\textstyle\frac{i}{2}}\,\sigma_{a}^{ik}v_{ik} \,,\qquad
|v|^2=v_av_a={\textstyle\frac{1}{2}}\,v^{ik}v_{ik}\,, \lb{Spin1}
\ee
where
$\sigma_{a}^{ik}=\epsilon^{ij}\sigma_{aj}{}^{k}$,
$\sigma_{a}{}_{ik}=\epsilon_{kj}\sigma_{ai}{}^{j}$ and $\sigma_{ai}{}^{k}$ are the standard
Pauli matrices.
Note that
$A^{ik}B_{ik}=2A_{a}B_{a}$, $\partial_a=i\,\sigma_{a}^{ik}\partial_{ik}$ and
$\partial^{ik}={\textstyle\frac{i}{2}}\,\sigma_{a}^{ik}\partial_a\,$.

In the vector notation, the constraint (\ref{const-pot-lin3}) for
$\mathscr{R}_{a}$ is rewritten as:
\begin{equation}\label{const-pot-lin3a}
\mathscr{R}_{a}={\textstyle\frac{1}{2}}\, \partial_{a}\mathscr{U}\,.
\end{equation}
The action (\ref{ac-bcom-lin1}) takes the form
\begin{equation}\label{ac-bcom-lin2}
S_{b} = \int dt \, \Big[ \dot{x}\dot{x} - {\textstyle\frac{1}{4}}\,(v_{a}+c_a)(v_{a}+c_a) -
\mathscr{A}_{a}\dot v_{a} + B\left( x- \mathscr{U}\right) \Big],
\end{equation}
while the constraints (\ref{const-pot-lin1}), (\ref{const-pot-lin2}) on the potentials
$\mathscr{A}_{a}$ and $\mathscr{U}$ are rewritten as:
\begin{equation}\label{const-pot-lin1-3}
\partial_a\partial_a\,\mathscr{U}=0\,,
\end{equation}
\begin{equation}\label{const-pot-lin2-3}
\partial_a\partial_a\,\mathscr{A}_{b}=0\,,\qquad
\partial_a\mathscr{A}_{a}=0\,,
\end{equation}
\begin{equation}\label{const-pot-lin3-3}
\mathscr{F}_{ab} := \partial_a\mathscr{A}_{b}-\partial_b\mathscr{A}_{a}=
-\epsilon_{abc}\partial_c\mathscr{U}\,.
\end{equation}
Note that the Laplace equation \p{const-pot-lin1-3} follows already from \p{const-pot-lin3-3} as the condition
ensuring the Bianchi identity for $\mathscr{F}_{ab}\,$.

Each solution of \p{const-pot-lin1-3} produces some static solution for the self-dual Maxwell potential
$\mathscr{A}_{a}\,$. For what follows, it is worth recalling the general multi-center solution. It is given by
\begin{equation}\label{pot-lin-general}
\mathscr{U}=\mathscr{U}_n:=g_0+\sum\limits_{s=1}^{n}\frac{g_s}{|\vec{v}-\vec{k}_s|}\,,
\end{equation}
where $g_0$ and $g_s$ are constants. Constant vectors $\vec{k}_s$ can be interpreted as  defining the
positions of the magnetic monopole charges. The constant $g_0$ specifies only asymptotic properties
of the potential (\ref{pot-lin-general}). Taking into account the constraint (\ref{constr-2}),
$x= \mathscr{U}$, the presence of non-zero constant $g_0$ in $\mathscr{U}$  amounts to the
trivial shift of the $x$ variable. Therefore, without loss of generality, we assume $g_0=0$ below.
For the potential (\ref{pot-lin-general}), the solution of the equations (\ref{const-pot-lin2-3}), (\ref{const-pot-lin3-3})
for the 3--vector potential $\vec{\mathscr{A}\,}=(\mathscr{A}_a)$ reads
\begin{equation}\label{vec-pot-lin-general}
\vec{\mathscr{A}\,}=\sum\limits_{s=1}^{n}\vec{\mathscr{A}\,}_s\,,\qquad
\vec{\mathscr{A}\,}_s=  g_s\,\frac{\vec{n}_s\times(\vec{v}-\vec{k}_s)}
{|\vec{v}-\vec{k}_s|\left(|\vec{n}_s||\vec{v}-\vec{k}_s| + \vec{n}_s(\vec{v}-\vec{k}_s)\right)}\,,
\end{equation}
where the non-physical 3--vectors $\vec{n}_s$ parametrize the Dirac string\footnote{Recall that the Dirac monopole field strength
does not display dependence on these variables.}.

Let us now perform the Hamiltonian analysis of the system with the action \p{ac-bcom-lin2}.

The relevant constraints are
\begin{equation}\label{constr-1}
\pi_{a}\equiv p_{a}+ \mathscr{A}_{a}\approx 0\,,
\end{equation}
\begin{equation}\label{constr-2}
h\equiv x- \mathscr{U}\approx 0\,,
\end{equation}
and the Hamiltonian reads:
\begin{equation}\label{Ham-b-lin}
H = {\textstyle\frac{1}{4}}\,p^2 +{\textstyle\frac{1}{4}}\,(v_{a}+c_a)(v_{a}+c_a) + \lambda_{a}\pi_{a}+Bh \,,
\end{equation}
where $\lambda_{a}$ and $B$ are the Lagrange multipliers. Poisson brackets
of the constraints (\ref{constr-1}), (\ref{constr-2}) are
\begin{equation}\label{PB-constr-l}
[\pi_a,\pi_b]_{{}_P}= -\mathscr{F}_{ab}\,,\qquad
[\pi_a,h]_{{}_P}= \partial_a\mathscr{U}\,,
\end{equation}
where  $\mathscr{F}_{ab}$ was defined in \p{const-pot-lin3-3}.
Determinant of the matrix of the right-hand sides of \p{PB-constr-l} is
$ (\partial_a\mathscr{U}\partial_a\mathscr{U})^2 \neq 0$ \footnote{The case when
$\partial_a\mathscr{U}=0$ and $\mathscr{U}=const$, $\mathscr{A}_a=0$, is trivial and so we
do not consider it.}. Hence, all four constraints (\ref{constr-1}), (\ref{constr-2}) are
second class.
The Dirac brackets corresponding to them are
\begin{equation}\label{DBconstr-lin}
[A,B]_{{}_D}=[A,B]_{{}_P}+ \frac{\epsilon_{abc}\partial_c\mathscr{U}}
{ \partial_p\mathscr{U}\partial_p\mathscr{U}}\,[A,\pi_a]_{{}_P}[\pi_b,B]_{{}_P}+
\frac{\partial_a\mathscr{U}} { \partial_p\mathscr{U}\partial_p\mathscr{U}}\, \Big(
[A,\pi_a]_{{}_P} [h,B]_{{}_P}-[A,h]_{{}_P} [\pi_a,B]_{{}_P} \Big)\,.
\end{equation}
For the phase variables, they yield
\begin{equation}\label{DBxp-lin}
[x,p]_{{}_D}=1\,,
\end{equation}
\begin{equation}\label{DBxv-lin}
[v_a,x]_{{}_D}=0\,,\qquad [v_a,p]_{{}_D}= \frac{\partial_a\mathscr{U}}
{ \partial_p\mathscr{U}\partial_p\mathscr{U}}\,,
\end{equation}
\begin{equation}\label{DBvv-lin}
[v_a,v_b]_{{}_D}= -\epsilon_{abc}\, \frac{\partial_c\mathscr{U}}
{ \partial_p\mathscr{U}\partial_p\mathscr{U}}\,.
\end{equation}

Now we have two independent physical phase variables ($x$ and $p$) and two independent spin variables,
hidden in $v_a$. Indeed, as follows from
the examples considered below, the constraint \p{constr-2} can be treated
as the equation defining a two-dimensional surface in the $\mathbb{R}^3$ manifold parametrized by the variables $v_a$.

\subsection{Nahm equations}
The Dirac brackets \p{DBxv-lin} and \p{DBvv-lin}  guarantee the fulfillment of the equations
\begin{equation}\label{Nahm-class}
[p,v_a]_{{}_D}={\textstyle\frac12}\,\epsilon_{abc}\,[v_b,v_c]_{{}_D}\,.
\end{equation}
Surprisingly, they are none other than a version of the famous Nahm equations \cite{Nahm}, with the Dirac bracket instead
of the commutators of the gauge group generators appearing in the original form of these equations.
We can define the ``genuine'' spinning variables $\ell_a$,
so that they decouple from the dynamical degrees of freedom $x$, $p$ with respect to the Dirac brackets,
i.e. $[\ell_a,x]_{{}_D}=[\ell_a,p]_{{}_D}=0$. Then $[p,v_a]_{{}_D}=-\frac{\partial}{\partial x}v_a := -v^\prime_a$
and the equations \p{Nahm-class} are rewritten as
\begin{equation}\label{Nahm-class1}
v^\prime_a=-{\textstyle\frac12}\,\epsilon_{abc}\,[v_b,v_c]_{{}_D}
\end{equation}
for $v_a{=}v_a(x,\ell_a)$. In this form they coincide with the generalized (the so called ``SDiff$(\Sigma_2)$'') Nahm equations,
as given, e.g., in \cite{Ward,GCPl,Dun}.

The mechanical model we are considering provides a dynamical realization of the close interconnection
between the three-dimensional Laplace equation and the SDiff$(\Sigma_2)$ Nahm equations established in \cite{Dun}.
Indeed, it is just the dim-3 Laplace equation \p{const-pot-lin1-3} which ensures the self-consistency
of the set of Dirac brackets  \p{DBxv-lin}, \p{DBvv-lin}, which in turn imply the Nahm
equations \p{Nahm-class1}.
Below we will see that the Nahm equations \p{Nahm-class} and their quantum counterpart ensure
the existence of the ${\cal N}{=}\,4$ supersymmetry in models with the $({\bf 3,4,1})$ spin multiplet,
both at the classical and the quantum levels.
\bigskip

In the next subsections we consider some simple examples of the models associated with the action \p{ac-bcom-lin1}.

\subsection{The one-monopole case}

Let us consider first the simplest one-monopole case, in which
\begin{equation}\label{pot-lin-1}
\mathscr{U}=\mathscr{U}_1:=\frac{g}{|\vec{v}-\vec{k}|}\,,\qquad
\mathscr{A}_a= g\,\frac{\epsilon_{abc}n_b ({v}_c-{k}_c)}
{|\vec{v}-\vec{k}|\left(|\vec{n}||\vec{v}-\vec{k}| + \vec{n}(\vec{v}-\vec{k})\right)}\,.
\end{equation}
The constant vector $\vec k$ can be absorbed into the redefinition of $v^a$, and in terms of
$\tilde{v}^a = v^a - k^a$ the potential $\mathscr{U}_1$ possesses manifest $SU(2) \sim SO(3)$ invariance.
The ``magnetic field'' $\nabla{\times}\vec{\mathscr{A}}$ points along the
radial direction, $\nabla{\times}\vec{\mathscr{A}}\sim \tilde{\vec{v}}$, i. e.
\begin{equation}\label{radial-mf}
\hat{v}_{a}\mathscr{A}_{a}=0\,.
\end{equation}
Obviously, in this case the whole Lagrangian \p{ac-bcom-lin2} is SU(2) invariant under the condition
\be
\vec k = - \vec c\,, \label{c-k}
\ee
where $c_a = {\textstyle\frac{i}{2}}\sigma_a^{ik}\,c_{ik}\,$.

According to \cite{IL}, the one-monopole potential
in \p{pot-lin-1} with $\vec n = \vec k/|\vec k|$ can be produced by  the following analytic superfield Lagrangian $\mathscr{L}^{(+2)}$
\begin{equation}\label{scof-ac}
\mathscr{L}^{(+2)} \sim \frac{L^{++}}{\left(
1+\sqrt{1 -k^{--}L^{++}}\right)\sqrt{1 - k^{--}L^{++}}}\,,
\end{equation}
which, besides ${\cal N}{=}4, d\,{=}1$ Poincar\'e supersymmetry, also exhibits ${\cal N}{=}4$ superconformal invariance
associated with the supergroup $D(2,1;\alpha)$ (it involves as a subgroup that SU(2) which provides
invariance of the bosonic action \p{ac-bcom-lin2}).
The total superfield action \p{A}, with
$\mathscr{L}({\mathscr{X}}) = -\frac{1}{2}{\mathscr{X}}^2\,$, exhibits the particular ${\cal N}{=}4$ superconformal SO(4$|$2)
invariance, provided the condition \p{c-k} is imposed \cite{FIL2}.

The constraint (\ref{constr-2}), i.e. $x= \mathscr{U}$, for the potential
(\ref{pot-lin-1}) becomes
\begin{equation}\label{constr-2a}
x={ g}/{|\vec{v}-\vec{k}|} \,.
\end{equation}
Now we introduce the new variables
\begin{equation}\label{ell-radial}
\ell_{a} =x\, (v_{a}-k_a) = g\frac{v_{a}-k_a}{|\vec{v}-\vec{k}|}\,.
\end{equation}
The constraint (\ref{constr-2}) (or, equivalently, (\ref{constr-2a})) then amounts to the condition
\begin{equation}\label{ell-radial-sq}
\ell_{a}\ell_{a} =  g^2\,.
\end{equation}
The Dirac brackets (\ref{DBxp-lin})-(\ref{DBvv-lin}) are rewritten as
\begin{equation}\label{DBxp-lin1a}
[x,p]_{{}_D}=1\,,
\end{equation}
\begin{equation}\label{DBxv-lin1a}
[\ell_a,x]_{{}_D}=0\,,\qquad [\ell_a,p]_{{}_D}= 0\,,
\end{equation}
\begin{equation}\label{DBvv-lin1a}
[\ell_a,\ell_b]_{{}_D}= \epsilon_{abc}\ell_c\,.
\end{equation}
Thus the variables $\ell_a$ parametrize a sphere $S^2$ with the radius $ g $ and
generate SU(2) group with respect to the Dirac brackets. The Nahm equations \p{Nahm-class1} are evidently satisfied by
$v_a = \frac{\ell_a}{x} + k_a$ as a consequence of \p{DBvv-lin1a}. After quantization, the variables $\ell_a$ are going
to parametrize a fuzzy sphere, with the relation \p{ell-radial-sq} becoming the SU(2)
Casimir condition for a fixed spin (``fuzziness'').

In the one-monopole case the Lagrangian in \p{ac-bcom-lin2} takes the form
\be
L^{bose} = \dot{x}\dot{x} - \frac{1}{4}\frac{|\vec{\ell}+x(\vec{k}+\vec{c})|^2}{x^2}
- \tilde{\mathscr{A}}_a\dot{\ell}_a\,, \lb{onemonBL}
\ee
where
$$
\tilde{\mathscr{A}}_a = g\,\frac{\epsilon_{abc}n_b \ell_c}
{|\vec{\ell}|\left(|\vec{n}||\vec{\ell}| + \vec{n}\vec{\ell}\right)}\,,
$$
i.e. it is a sum of the conformal mechanics Laqrangian and SU(2) WZ term. Respectively, the Hamiltonian
(\ref{Ham-b-lin}) reads
\begin{equation}\label{Ham-b-lin1}
H = \frac{1}{4}\left(p^2 + \frac{|\vec{\ell}+x(\vec{k}+\vec{c})|^2}{x^2}\right) \,.
\end{equation}
The requirement of preservation of the vector $\vec{\ell}$, $[H,\ell_a]_{{}_D}=0$, leads  just to the
condition \p{c-k}.
Then, the bosonic Hamiltonian finally becomes
\begin{equation}\label{Ham-b-lin2}
H = \frac{1}{4}\left(p^2 + \frac{\ell_a\ell_a}{x^2}\right) =  \frac{1}{4}\left(p^2 + \frac{g^2}{x^2}\right).
\end{equation}

The one-monopole system under consideration can be quantized in a few different ways,
depending on the quantum realization of the spin variables $\ell_a$.

Let us firstly consider one possible  quantization scheme,
which explicitly takes into account the properties of fuzzy sphere.
After quantization variables $\ell_a$ become operators
\begin{equation}\label{l-L}
\ell_a   \qquad\rightarrow\qquad  \hat\ell_a \,.
\end{equation}
The commutation relations of $\hat\ell_a$ are determined by the Dirac brackets \p{DBvv-lin1a}
\footnote{We quantize by replacing $[ A, B\}_{{}_D}{=}\, C$ $\rightarrow$ $[\hat A,\hat B\}{=}\,i\hbar\, \hat C$
for basic variables and explicitly keep the Planck constant $\hbar$ in all quantum expressions, having in mind that in
section 4 the quantum analogs of some classical quantities will be sought for as a power series in $\hbar\,$.}
and form the $su(2)$ algebra
\begin{equation}\label{LL-lin1a}
[\hat\ell_a,\hat\ell_b]= i\hbar\,\epsilon_{abc}\hat\ell_c\,.
\end{equation}

The constraint \p{ell-radial-sq} should hold for the operators $\hat\ell_a$ in the strong sense,
because we quantize Dirac brackets.
On the other hand, the quantity $\hat\ell_a\hat\ell_a$ is the Casimir operator for $su(2)$.
Therefore, for unitary representations it must be equal to
\begin{equation}\label{LL-const}
\hat\ell_{a}\hat\ell_{a} =\hbar^2 n(n+1)
\end{equation}
where $n$ is a non-negative half-integer or integer number, i.e.
$2n\in \mathbb{N}$. Thus, in the process of quantization
the classical constant $g^2$ present in the constraint \p{ell-radial-sq} should be substituted as
\begin{equation}\label{g-n}
g^2 \qquad\rightarrow\qquad  \hbar^2 n(n+1),
\end{equation}
i.e. it gets quantized.

Then, for $\hat\ell_a$ we can use the standard realization by
$(2n+1){\times}(2n+1)$ matrices. As a result,
the wave function has $(2n+1)$ components
and describes a non--relativistic spin $n$ conformal particle.
The corresponding Hamiltonian is \p{Ham-b-lin2}, in which the replacements
\p{l-L}, \p{g-n} are done. The full set of the conformal symmetry generators
will be given below, while considering the supersymmetric case.

Now we apply to a different quantization method which will be also used in the two-center case.
Its main idea is to describe the spinning sector by two independent variables.
One can define the variables
\begin{equation}\label{def-ell3-phi-1}
\ell_3\qquad \mbox{and} \qquad
\varphi:= \arctan\left(\frac{\ell_2}{\ell_1}\right),
\end{equation}
which still represent the two-sphere and have the canonical Dirac bracket
\begin{equation}\label{DB-ell-phi-1}
[\varphi,\ell_3]_{{}_D}= 1\,.
\end{equation}
Being rewritten through the variables $\varphi$ and $\ell_3$, the WZ term in \p{onemonBL}
takes the very simple form
\begin{equation}\label{WZ-bcom-lin2}
-\int \tilde{\mathscr{A}}_{a}\,d \ell_{a} =\int \ell_3\,d \varphi\,.
\end{equation}

It is easy to show that the WZ action \p{WZ-bcom-lin2} is invariant (up to a total time derivative in the integrand)
under the SU(2) transformations realized as
a particular subgroup of the general group of symplectic diffeomorphism of the surface $(\ell_3, \varphi)$:
\bea
&&\delta \varphi = \frac{\partial f(\varphi, \ell_3)}{\partial \ell_3}\,, \quad \delta \ell_3
= -\frac{\partial f(\varphi, \ell_3)}{\partial \varphi}\,, \lb{l3phi} \\
&& f(\varphi, \ell_3) = a_3 \ell_3 + \sqrt{g^2 - \ell_3^2}\left(a_1\cos \varphi + a_2 \sin \varphi\right),\lb{l3phi1}
\eea
where $a_{1,2,3}$ are properly normalized parameters of SU(2).\footnote{The WZ term \p{WZ-bcom-lin2} is invariant
(up to a total $t$-derivative) under the full symplectic diffeomorphism group \cite{FRS,HTown}. However, this invariance is broken
down to SU(2) in the full Lagrangian with fermions.} This SU(2) is a symmetry of the classical theory, so it
is natural to require it to be preserved at the quantum level too. The classical SU(2) generators can be constructed as
\begin{equation}\label{ell-sph}
\ell_1= \sqrt{g^2 - \ell_3^2}\,\cos\varphi\,,\qquad \ell_2=\sqrt{g^2 - \ell_3^2}\,\sin\varphi\,,\qquad\ell_3\,,
\ee
or
\be
\ell_1=g\sin\vartheta\cos\varphi\,,\qquad \ell_2=g\sin\vartheta\sin\varphi\,,\qquad\ell_3=g\cos\vartheta\,,
\end{equation}
where $\vartheta$ is the second (azimuthal) angle on the sphere. It is easy to check that these quantities form
the classical SU(2) algebra \p{DBvv-lin1a} with respect to the Dirac bracket \p{DB-ell-phi-1} (and in fact generate
the transformations \p{l3phi}).
However, the direct passing
to the quantum case via replacing the Dirac brackets by commutators can be plagued by the ordering ambiguities. It is convenient
to pass to the complex variable
\begin{equation}\label{def-z-1}
z:= e^{\,i\varphi}\cot\left(\vartheta/2\right)\,,
\end{equation}
\begin{equation}\label{DB-zz-1}
[z,\bar z]_{{}_D}= \frac{i}{2g}\,(1+z\bar z)^{2}\,.
\end{equation}
In terms of  $z$ and $\bar z$ the generators \p{ell-sph} take the form
\begin{eqnarray}\nonumber
\ell_+=\ell_1+i\ell_2&=& \frac{2gz}{1+z\bar z}=2gz-z^2\,\frac{2g\bar z}{1+z\bar z}\,,\\ [5pt]\label{ell-z}
\ell_-=\ell_1-i\ell_2&=& \frac{2g\bar z}{1+z\bar z}\,,\\ [5pt]\nonumber
\ell_3&=& -g\,\frac{1-z\bar z}{1+z\bar z}=z\,\frac{2g\bar z}{1+z\bar z}-g\,.
\end{eqnarray}
Since
\begin{equation}\label{DBr-zpz-1}
\left[z,\frac{2g\bar z}{i(1+z\bar z)}\right]_{{}_D}= 1\,,
\end{equation}
the quantum counterpart of $2g\bar z/(1+z\bar z)$ plays the role of $\partial_z$ in the holomorphic representation, i.e.
$$
\frac{2g\bar z}{(1+z\bar z)} \rightarrow \frac{\partial}{\partial z}
$$
after quantization.
Then the holomorphic quantum realization of the SU(2) generators \p{ell-z} is as follows \cite{Per,FRS,IM}
\begin{equation}\label{hat-ell-1}
\begin{array}{rcl}
\hat\ell_1&=& \hbar\left[\frac12\,(1-z^2)\partial_z+gz\right],\\ [5pt]
\hat\ell_2&=& \hbar\left[\frac{i}{2}\,(1+z^2)\partial_z-igz\right],\\ [5pt]
\hat\ell_3&=& \hbar\left(z\partial_z-g\right),
\end{array}
\end{equation}
where, for coherence, we again restored the Planck constant. The possible ordering ambiguity can always be absorbed
into a redefinition of the parameter $g$.

At fixed $g$, the Hilbert space is spanned by $2g+1$ basis wave functions $1$, $z$, ..., $z^{2g}$
with the inner product defined as \cite{Per,FRS,IM}
\begin{equation}\label{DB-zpz-1}
\langle\Psi,\Phi\rangle =\frac{2g+1}{2\pi i}\,\int_{S^2} \frac{dzd\bar z}{(1+z\bar z)^{2g+2}}\,\bar\Psi(\bar z)\Phi(z)\,.
\end{equation}
The norms  $|\Phi|^2 = \langle\Phi,\Phi\rangle$ are finite, and this property amounts to saying that $\Phi$ and $\Psi$
are square-integrable (and hence well defined) functions on $\mathbb{CP}^{1} \sim S^2$.
In the realization \p{hat-ell-1}, $\hat\ell_a\hat\ell_a=\hbar^2g(g+1)\,$, i.e. the basis functions span an irreducible
spin $g$ multiplet of the group SU(2).\footnote{In fact, requiring the norm of $\Phi(z)$ to be convergent with respect to the
inner product \p{DB-zpz-1} already restricts $2g$ to be integer and $\Phi(z)$ to be a polynomial in $z$ of degree $2g$.}
Thus in this quantization scheme the original constant $g$ is identified
with the spin quantum number $n \in \mathbb{Z}, \mathbb{Z} + \frac{1}{2}$, as opposed to
the quantization formula \p{g-n} of the previously employed method.\footnote{An equivalent approach
is the Gupta-Bleuler (or ``geometric'')  quantization on the two-sphere (see e.g. \cite{IM}).}

\subsection{The two-monopole case}

This case corresponds to keeping two terms in the general solution \p{pot-lin-general} (with $g_0=0$)
of the Laplace equation for the function $\mathscr{U}$.
Without loss of generality, we can place the singularity point on the axis $z$,
other possible choices are obtained by proper rotations and shift of the coordinate system.
Thus, the two--monopole potential can be chosen in the form
\begin{equation}\label{pot-lin-2a}
\mathscr{U} \equiv \mathscr{U}_2:=\frac{g_1}{|\vec{v}-\vec{k}_1|}+\frac{g_2}{|\vec{v}-\vec{k}_2|}\,,
\end{equation}
where
\begin{equation}\label{def-v+-}
 \vec{k}_1=(0,0,k_1)\,,\qquad \vec{k}_2=(0,0,k_2)\,.
\end{equation}
The corresponding analytic superfield Lagrangian $\mathscr{L}^{(+2)}$ is a sum of two Lagrangians \p{scof-ac},  with
the parameters $k_1^{ik}$ and $k_2^{ik}\,$. It possesses only U(1) internal symmetry and does not exhibit
superconformal symmetry (unless $k_1^{ik} = k_2^{ik}\,$ is assumed, that would take us back to the one-monopole case).

To quantize the system, we can proceed by analogy with the one-monopole case. We should pass from the variables $v_a$
to some new variables, such that their Dirac brackets are maximally simple. One of these variables is
the ``dynamical'' degree of freedom $x\,$, whereas two residual degrees of freedom are spin ones.
In these new variables, the spin sector should decouple from the ``dynamical'' sector $(x,p)$.
This separation of the true dynamical degrees of freedom from the ``semi-dynamical''
spin degrees of freedom is a necessary step in performing quantization of the relevant Dirac brackets.

Using the basic brackets  (\ref{DBxp-lin})-(\ref{DBvv-lin}), it is direct to check that the quantity
\begin{equation}\label{ell3-def}
\ell_3:=\frac{g_1(v_3-k_1)}{|\vec{v}-\vec{k}_1|}\,+\,\frac{g_2(v_3-k_2)}{|\vec{v}-\vec{k}_2|}
\end{equation}
commutes with the variables of the dynamical sector:
\begin{equation}\label{ell3-br}
[\ell_3,p\,]_{{}_D}=[\ell_3,x\,]_{{}_D}=0\,.
\end{equation}
The second semi-dynamical spin degree of freedom is the polar angle coordinate of $\vec{v}$
\begin{equation}\label{phi-def}
\varphi:= \arctan\left(\frac{v_2}{v_1}\right)\,,
\end{equation}
\begin{equation}\label{phi-br}
[\varphi,p\,]_{{}_D}=[\varphi,x\,]_{{}_D}=0 \,.
\end{equation}
Moreover, the variables (\ref{ell3-def}) and (\ref{phi-def}) are conjugate to each other with respect to the Dirac bracket:
\begin{equation}\label{phi-ell3}
[\varphi,\ell_3\,]_{{}_D}=1 \,.
\end{equation}

Thus, the phase space of the model decouples in the two sectors: one is the dynamical sector spanned by the pair $(x,p)\,$,
whereas the pair $(\varphi,\ell_3)$ defines the semi-dynamical spin sector. Similarly to the one-center case,
the WZ term in the action (\ref{ac-bcom-lin2}) have the form $\int dt\,\ell_3\,\dot \varphi$ in
terms of the variables $\varphi$ and $\ell_3$.

Note that the variable $\ell_3$ has the clear physical meaning: it is just the N\"{o}ther conserved charge
for the O(2) phase transformations
\begin{equation}\label{tr-o2}
\delta v_1=\alpha\, v_2\,, \qquad \delta v_2=-\alpha\, v_1\,, \qquad \delta v_3=0\,,
\end{equation}
when $c_a$ points along the third axis, $c_a=(0,0,c)\,$. Namely this case, when the vector $\vec c$ is collinear to the vectors
$\vec k_{1,2}\,$ and
\begin{equation}\label{DB-Hl13}
\qquad [\ell_{3},H]_{{}_D} = 0\,,
\end{equation}
will be considered below.

The inverse relations $v_a=v_a(x,\ell_3,\varphi)$ are more
complicated, and in what follows they will not be used in the closed
form. Rather, we will use the relations (\ref{pot-lin-2a}),
(\ref{ell3-def}), (\ref{phi-def}) to express $|\vec{v}|$ and $v_3$
in terms of $x{=}\mathscr{U}$ and $\ell_3$ via some recurrence
procedure with respect to a small parameter. Let us illustrate this
on two cases. \vspace{0.3cm}

{\bf i)}\, It is the case when the location of the second pole tends to infinity. Using the notation
$$
g_1  = g\,,\qquad  \vec{k}_1=\vec{k}\,,\qquad  \vec{k}_2=\vec{k}+ \vec{k}/\varepsilon\,,
$$
we find that in the limit $\varepsilon\to \,0$,  when
$$
E=\frac{\varepsilon^2g_2}{k^2}\ll 1\,,\qquad g_0=\frac{\varepsilon g_2}{k}\ll 1\,,
$$
the following relation holds
\begin{equation}\label{pot-lin-2a-1}
x=\mathscr{U}_2 \equiv \mathscr{U}^\prime_2:=\frac{g}{|\vec{v}-\vec{k}|}+2E(v_3-k)+g_0\,.
\end{equation}
As was mentioned in section 3.1, a non-zero constant $g_0$ in the general two-center potential (\ref{pot-lin-general}) can always
be absorbed into a constant shift $x\to x^\prime=x-g_0\,$. The same can be done at each step of considering the $\varepsilon$
expansion of the potential: an additive constant in any order can be removed by a similar shift of $x\,$. In particular, in \p{pot-lin-2a-1}
one can put $g_0 = 0\,$. Thus the only essential small parameter in the $\varepsilon$- expansion will be $E\,$.
We will omit the primes on $x\,$, hoping that this will not give rise to any confusion.

Thus, in the limit of small $\varepsilon$ the potential (\ref{pot-lin-2a-1}) is reduced to the sum of the one-monopole potential  (\ref{pot-lin-1})
and a constant ``electric background field potential'' $\sim \vec{E}(\vec{v}-\vec{k})$.
The U(1) N\"{o}ther charge in the case of the potential (\ref{pot-lin-2a-1}) is given by the expression
\begin{equation}\label{ell3-lin-2a-1}
\ell_3=\frac{g(v_3-k)}{|\vec{v}-\vec{k}|}\,+\,E\left(|\vec{v}-\vec{k}|^2-(v_3-k)^2\right).
\end{equation}
Now we can invert the expressions (\ref{pot-lin-2a-1}), (\ref{ell3-lin-2a-1}) to represent $|\vec v|$ and $v_3$
as a power series in the small parameter $E$
\begin{eqnarray}\label{v3-lin-2a-1}
v_3-k &=&
\frac{\ell_3}{x} +E\,\frac{3\ell_3^2-g^2}{x^3} +{\cal O}(E^2)\,, \\
\sqrt{(v_1)^2+(v_2)^2} &=&
\frac{\sqrt{g^2-\ell_3^2}}{x}\left(1 +E\,\frac{3\ell_3}{x^2}\right) + {\cal O}(E^2) \,.\label{v12-lin-2a-1}
\end{eqnarray}
In the expansions (\ref{v3-lin-2a-1}), (\ref{v12-lin-2a-1}), the higher-order terms are determined
by the lower-order ones by the recurrence procedure, with taking account of the next orders in \p{pot-lin-2a-1} and \p{ell3-lin-2a-1}, etc.
\vspace{0.3cm}

{\bf ii)}\, The second limiting case corresponds to the situation when the distance between poles tends to zero:
$$
g_1  = g_2  =g/2\,,\qquad  \vec{k}_1=\vec{k}\,,\qquad  \vec{k}_2=\vec{k}+ \varepsilon\vec{k}\,.
$$
In the limit $\varepsilon\to 0$ we obtain
\begin{equation}\label{pot-lin-2a-2}
x=\mathscr{U}_2 \equiv \mathscr{U}^{\prime\prime}_2:=\frac{g}{|\vec{v}-\vec{k}|}+\frac{d(v_3-k)}{|\vec{v}-\vec{k}|^3}\,,
\end{equation}
where
$$
d=\varepsilon kg\ll 1\,.
$$
The potential (\ref{pot-lin-2a-2}) is the sum of the one-monopole potential  (\ref{pot-lin-1})
and the ``electric dipolar potential'' $\sim \vec{\,d}(\vec{v}-\vec{k})/|\vec{v}-\vec{k}|^3$.
The generator of the U(1) symmetry transformations in this case is expressed as
\begin{equation}\label{ell3-lin-2a-2}
\ell_3=\frac{g(v_3-k)}{|\vec{v}-\vec{k}|}\,-\,d\,\frac{|\vec{v}-\vec{k}|^2-(v_3-k)^2}{|\vec{v}-\vec{k}|^3}\,.
\end{equation}
The expressions inverse to (\ref{pot-lin-2a-2}), (\ref{ell3-lin-2a-2}) are represented in the form of a series
in the small parameter $d$  as
\begin{eqnarray}\label{v3-lin-2a-2}
v_3-k &=&
\frac{\ell_3}{x} +d\,\frac{1}{g} -d^2\,\frac{x\ell_3}{g^4}+d^3\,\frac{x^2(3\ell_3^2-g^2)}{g^7} +{\cal O}(d^4)\,,\\
\sqrt{(v_1)^2+(v_2)^2} &=&
\frac{\sqrt{g^2-\ell_3^2}}{x} \left(1+d^2\,\frac{x^2}{2g^4}-d^3\,\frac{2x^3\ell_3}{g^7}\right) +{\cal O}(d^4) \,.\label{v12-lin-2a-2}
\end{eqnarray}
Like in the previous case, higher-order terms in the expansions (\ref{v3-lin-2a-2}), (\ref{v12-lin-2a-2})
are determined by the lower-order ones by recurrence, with taking into account the next orders in $d$ in \p{pot-lin-2a-2} and \p{ell3-lin-2a-2}.

Choosing, as in one-monopole case, $\vec{c}=-\vec{k}$ in the Hamiltonian (\ref{Ham-b-lin}) and using there the expression
(\ref{v3-lin-2a-1})-(\ref{v12-lin-2a-1}) or (\ref{v3-lin-2a-2})-(\ref{v12-lin-2a-2})
we obtain the Hamiltonian in terms of the variables $p$, $x$ and $\ell_3$ as a series in the relevant small parameters.

We observe that in the two-monopole case the expansion for $v_3$ (see (\ref{v3-lin-2a-1}) and (\ref{v3-lin-2a-2}))
contains not only the linear terms in $\ell_3$ as in the one-monopoly case, but also terms of the higher-order in
this variable. The presence of these higher-order terms will lead  to the essential modification
of the quantization procedure in the full-fledged supersymmetric case.

In the expansions (\ref{v3-lin-2a-1}) and (\ref{v3-lin-2a-2}), we will leave only the first nontrivial corrections
quadratic in $\ell_3$. Note that in the considered case such terms appear in the third order in $d$,
while in the previous case already in the first order in $E\,$. This is the reason why we restrict our attention
to the first and third orders in the relevant $\varepsilon$ expansions.
\vspace{0.3cm}

The commutator of the quantum operators $\hat\varphi$ and $\hat\ell_3$
corresponding to the classical Dirac bracket (\ref{phi-ell3}) is
\begin{equation}\label{phi-ell3-q}
[\hat\varphi,\hat\ell_3\,] =i \hbar\,.
\end{equation}
Obviously, we can choose the natural angular-momentum representation for these operators
\begin{equation}\label{rep-phi-ell3}
\hat\ell_3 =-i\hbar\,\partial/\partial\varphi\,,\qquad\quad\hat\varphi=\varphi\,,\qquad 0\leq\varphi\leq 2\pi \,.
\end{equation}
The wave function is then represented by the Fourier series
\begin{equation}\label{wf-2}
\Psi(x,\varphi) =\sum_{n=-\infty}^{\infty}e^{in\varphi}\psi_n(x)\,.
\end{equation}
The component $d=1$ fields $\psi_n(x)$ describe the states with a fixed value of the U(1) ``momentum'' generator $\hat\ell_3\,$.
We can keep only one component in the expansion (\ref{wf-2}), thus choosing some fixed  irreducible representation
of U(1). This is sufficient for constructing a quantum system which respects the same U(1) symmetry as
in the classical case. Alternatively, we could pick up a many-component wave function on which
more general SU(2) group is realized, in the same way as in one-monopole case (\ref{def-z-1})-(\ref{hat-ell-1}).
In this case we will obtain a quantum system with the spinning sector still represented by fuzzy sphere.
Such a system will be considered in the next subsection.

\subsection{A special multi-monopole case}

Here we consider the potential of the form
\begin{equation}\label{pot-tilde}
{\mathscr{U}}\equiv \tilde{\mathscr{U}}=\frac{g}{k}\,\,{\rm{arcoth}}\left( \frac{|\vec{v}+\vec{k}|+|\vec{v}-\vec{k}|}{2k}\right),
\end{equation}
where $\vec{k}=(0,0,k)$.
By Euler homogeneity operator $v_a\partial_a$
the potential (\ref{pot-tilde}) reproduces  the standard two-center potential (\ref{pot-lin-2a}) \cite{Dun}
(see also \cite{GORV,Gib}):
\begin{equation}\label{pot-lin-2}
{\mathscr{U}}_2= -v_a\partial_a \tilde{\mathscr{U}}= \frac{g}{2}\left(\frac{1}{|\vec{v}+\vec{k}|}+\frac{1}{|\vec{v}-\vec{k}|}\right)\,.
\end{equation}
Besides the two poles at $\vec{v} = \pm \vec{k}$, the potential \p{pot-tilde} possesses the third pole at $\vec{v} = 0\,$.

Choosing the potential in the form  (\ref{pot-tilde}) makes it possible to define
the spin variables in a way similar to the one-monopole case and, what is most important,
to quantize the corresponding full-fledged  supersymmetric model by analogy with that associated with
the one-monopole potential. We should pass from the variables $v_a$
to the new triplet of variables  $\ell_a$, such that their Dirac brackets with $p$ vanish, $[\ell_a,p\,]_{{}_D}=0\,$.
In terms of these new variables, the spinning sector decouples from the ``dynamical'' sector $(x,p)$.
This separation of the true dynamical degrees of freedom from the ``semi-dynamical''
spin degrees of freedom is a necessary step in quantization of the relevant Dirac brackets.

We split $v_a$ into the ``radial variable'' $x$ and the spin ones $\ell_a$ by the following relations
\begin{equation}\label{new-prop-1}
v_1=f_1(x)\,\ell_{1}\,, \qquad v_2=f_2(x)\,\ell_{2}\,, \qquad v_3=f_3(x)\,\ell_{3}\,,
\end{equation}
where
\begin{equation}\label{sol-f123}
f_1=f_2=\frac{k}{g\,\sinh(k x/g)}\,,
\qquad
f_3=\frac{k}{g}\,{{\rm{coth}}(k x/g)}\,.
\end{equation}
The functions $f_a$ thus defined satisfy the Euler equations
\footnote{Note that the functions $f_a$ are particle case of elliptic functions since
$\rm{cs}(z;m)=\rm{ds}(z;m)=1/\rm{sinh}(z)$, $\rm{ns}(z;m)=\rm{coth}(z)$ when $|m|=1$.}
\begin{equation}\label{new-eq-1}
f_1^\prime=-f_2f_3\,, \qquad f_2^\prime=-f_1f_3\,, \qquad f_3^\prime=-f_1f_2\,.
\end{equation}

The Dirac brackets  (\ref{DBxp-lin})-(\ref{DBvv-lin}), being rewritten through the new variables, take the form
\begin{equation}\label{DBxp-lin1a-1}
[{x},{p}]_{{}_D}=1\,, \qquad
[\ell_a,{x}]_{{}_D}=0\,,\qquad [\ell_a,{p}]_{{}_D}= 0\,,\qquad
[\ell_a,\ell_b]_{{}_D}= \epsilon_{abc}\ell_c\,,
\end{equation}
whereas the constraint (\ref{constr-2}) becomes the 2-sphere condition
\begin{equation}\label{ell-radial-sq-1}
\ell_{a}\ell_{a} =  g^2\,.
\end{equation}
The Hamiltonian (\ref{Ham-b-lin}) is rewritten as
\begin{equation}\label{Ham-b-lin2a}
H = \frac{1}{4}\,p^2 +
\frac{k^2}{4g^2}\,{\rm{sinh}}^{-2}(\frac{k\hat x}{g})
\left[ (\hat\ell_1)^2+(\hat\ell_2)^2+
\left({\rm{cosh}}(\frac{k\hat x}{g})\,\hat\ell_3+\frac{cg}{k}\,{\rm{sinh}}(\frac{k\hat x}{g})\right)^2\right] ,
\end{equation}
where we take $\vec{c}=(0,0,c)$ in order to obey the requirement that $\ell_{3}$
has the vanishing Dirac bracket with the Hamiltonian and hence generates U(1) symmetry of the system.
Thus, we obtain the Hamiltonian system which gives an opportunity to proceed  to the quantization
along the same line as in the one-monopole case.
Note that the relations (\ref{new-prop-1}) reproduce the one-monopole solution (\ref{ell-radial}) in the limit $k\to 0$.
In addition, the Nahm equations (\ref{Nahm-class1}) are satisfied with the Dirac brackets (\ref{DBxp-lin1a-1})
and the Euler equations (\ref{new-eq-1}).

As opposed to the one-monopole model of section\,3.1, where all the components of $\ell_a$, $a=1,2,3\,$, commute
with $x$, $p$ and Hamiltonian, the considered case is quite analogous to the model based on
the standard two-center potential (\ref{pot-lin-2a}), in which only $\ell_3$ can have vanishing Dirac bracket with $H$,
\begin{equation}\label{DB-Hl31}
\qquad [\ell_{3},H]_{{}_D} = 0\,,
\end{equation}
whereas
\begin{equation}\label{DB-Hl121}
[\ell_{1},H]_{{}_D} = -\alpha\, \ell_{2}, \,
\qquad [\ell_{2},H]_{{}_D} = \alpha\, \ell_{1}, \, \qquad
\alpha:= \frac{k^2\ell_3}{2g^2} \,.
\end{equation}
Therefore, only the third component $\ell_{3}$ is the `true' conserved quantity in the considered case,
while $\ell_{1}$ and $\ell_{2}$ are not.
However, the Dirac brackets (\ref{DB-Hl121}) show that the evolution of the variables $\ell_{1,2}$ with time
amounts to their U(1) rotation with some field-dependent parameter. Then the transformed vector
\begin{equation}\label{var-b}
\tilde{\ell}_a \simeq \ell_a+\delta t\,\dot\ell_a =\ell_a+ \delta t \, [\ell_a, H]_{{}_D}
\end{equation}
satisfies the same basic relations (\ref{DBxp-lin1a-1})
of the deformed fuzzy sphere (with the replacement $\ell_a\to\tilde{\ell}_a$). Thus, the (deformed) fuzzy sphere is preserved
under the dynamical evolution of the system and we can still use the standard quantum realization
of its coordinates by the $(2n+1){\times}(2n+1)$ matrices, as described at the end of the section\,3.1.
\vspace{0.5cm}

To summarize, we succeeded in performing the quantization of the bosonic limit of our ${\cal N}{=}4$ supersymmetric mechanics model in a closed form,
not only in the superconformally-invariant one-monopole case,
but also in the special two-monopole case, which preserves only $d=1$ Poincar\'e supersymmetry.
While the first case was already considered in \cite{FIL2}, based upon the ({\bf 4,4,0}) spin multiplet,
the second option was not addressed before. We managed to quantize due to passing to the proper spin variables
with the clear geometric meaning, based upon the requirement that the spin sector decouples from the
physical variables $x$ and $p$.  Although the new spin variables are related to the original variables by
the rather involved nonlinear transformations, they have simple Dirac brackets and, therefore,
admit a rather straightforward quantum realization. We fixed the redundancy in the definition
of the new spin variables by requiring them to form $su(2)$ algebra as in the one-monopole case,
and so to parametrize a fuzzy sphere.
This particular quantization scheme is distinguished in that the spin variables have a simple matrix quantum representation.
As distinct from the one-monopole case, this $su(2)$ is not a genuine symmetry of the system.
Only one of the spin variables, just $\ell_3$,  commutes with the Hamiltonian and so generates the genuine U(1) symmetry.
Nevertheless, the SU(2) algebra and its Casimir operator at the fixed fuzziness are preserved
under the time evolution.

\setcounter{equation}0

\section{Turning on supersymmetry}

As in the previous section we consider the particular case with $\mathscr{L}\vert=-\frac{1}{2}\,x^2$.
Then the total action (\ref{A-com-lin-1})  takes the following form
\begin{equation}\label{Act-full}
S=S_{b} +S_{f}\,,
\end{equation}
where the pure bosonic part $S_{b}$ was defined in  (\ref{ac-bcom-lin2}), whereas the terms with
fermionic fields are
\begin{equation}\label{A-com-ferm}
S_{f} =
i \left(\dot{\bar\chi}{}^k \chi_k - {\bar\chi}^k \dot{\chi}_k\right)
- {\textstyle\frac{i}{2}} \, (\mathscr{R}^{-1})^{ik} \chi_i \bar\chi_k \,.
\end{equation}

The action  (\ref{Act-full}) is invariant under the ${\cal N}{=}4$ supersymmetry transformations
\begin{equation}\label{supertransform}
\begin{array}{c}
\delta x= -\varepsilon_{i}\chi^{i} + \bar{\varepsilon}^{i}\bar{\chi}_{i}\,, \\ \\
\delta \chi^{i} =  i\,  \dot{x} \bar{\varepsilon}^{i} -{\textstyle\frac{i}{2}} \left(v^{ik}+c^{ik}\right) \bar{\varepsilon}_{k}\,,
\qquad
\delta \bar{\chi}_{i} = - i\, \dot{x} \varepsilon_{i} -{\textstyle\frac{i}{2}} \left( v_{ik}+c_{ik}\right)\varepsilon^{k}\, , \\ \\
\delta v^{ij} = -(\mathscr{R}^{-1})^{\,k(i} \left[\varepsilon^{j)} \chi_{k} + \bar{\varepsilon}^{j)} \bar{\chi}_{k}\right], \qquad
\delta B = -{\textstyle\frac{1}{2}} \,\frac{d}{dt} \Big[(\mathscr{R}^{-1})_{\,ik} (\varepsilon^{i} \chi^{k} + \bar{\varepsilon}^{i}
\bar{\chi}^{k})\Big],
\end{array}
\end{equation}
where $\varepsilon_i, \bar\varepsilon{}^i$ are the Grassmann parameters.
The corresponding supercharges are
\begin{equation}\label{superQ}
Q^{i} = p\, \chi^{i} +  \left(v^{ik} + c^{ik}\right)  \chi_{k} -{\textstyle\frac{1}{2}}
\left(x - \mathscr{U}^{\phantom{i}}\!\right) (\mathscr{R}^{-1})^{\,ik}\chi_{k}\,,
\end{equation}
\begin{equation}\label{superQbar}
\bar{Q}_{i} = p\, \bar{\chi}_{i} -  \left(v_{ik} + c_{ik}^{\phantom{i}}\right) \bar{\chi}^{k}
+ {\textstyle\frac{1}{2}}
\left(x - \mathscr{U}^{\phantom{i}}\!\right) (\mathscr{R}^{-1})_{\,ik} \bar{\chi}^{k}\,.
\end{equation}
The full Hamiltonian has the form
\begin{equation}\label{H-su}
H = {\textstyle\frac{1}{4}} \,p^{2} + {\textstyle\frac{1}{8}}
\left(v^{ik}+c^{ik}\right)\left(v^{\phantom{i}}_{ik}+c_{ik}\right)
+{\textstyle\frac{i}{2}} \, (\mathscr{R}^{-1})^{ik} \chi_i \bar\chi_k
- B\left(x- \mathscr{U}^{\phantom{i}}\! \right).
\end{equation}

Bosonic variables are subjected to the second class constraints (\ref{constr-1}), (\ref{constr-2}) and, as before,
we use Dirac brackets (\ref{DBconstr-lin}) for them. As a result,
the last terms in the supercharges and the Hamiltonian (\ref{superQ})-(\ref{H-su}) vanish,
and these quantities are finally expressed as
\begin{equation}\label{superQ-v}
Q^{i} = p\, \chi^{i} +  i\left(v_{a} + c_{a}\right)\sigma_a^{ik}  \chi_{k} \,,
\qquad
\bar{Q}_{i} = p\, \bar{\chi}_{i} -  i\left(v_{a} + c_{a}^{\phantom{i}}\right)\sigma_{a\,ik} \bar{\chi}^{k}
\,,
\end{equation}
\begin{equation}\label{H-su-v}
H = {\textstyle\frac{1}{4}} \,p^{2} + {\textstyle\frac{1}{4}}
\left(v^{a}+c^{a}\right)\left(v^{\phantom{i}}_{a}+c_{a}\right)
- \chi_i \sigma_a^{ik}\bar\chi_k\, \partial_{a} \mathscr{U} / (\partial_{p} \mathscr{U} \partial_{p} \mathscr{U})
\,,
\end{equation}
where we passed to the vector notations. Dirac brackets of the bosonic variables are the same as in (\ref{DBxp-lin})-(\ref{DBvv-lin}).
Fermionic variables $\chi^{i}$ and $\bar{\chi}_{i}$ have the following non-vanishing Dirac brackets:
\begin{equation}\label{PB-constr-la}
\{\chi^{i},\bar{\chi}_{k}\}_{{}_D}= -{\textstyle\frac{i}{2}}\,\delta^{\,i}_k\,.
\end{equation}
We checked that the operators (\ref{superQ-v}), (\ref{H-su-v}) form ${\cal N}{=}\,4$ supersymmetry algebra:
\begin{equation}\label{susy-alg}
\{Q^{i},\bar{Q}_{k}\}_{{}_D}= -2i\,\delta^{\,i}_k H\,,\qquad \{Q^{i},Q^{k}\}_{{}_D}=[Q^{i},H]_{{}_D}=0 \,.
\end{equation}

An important property of the supercharges (\ref{superQ-v})  is that
the basic relations of the supersymmetry \p{susy-alg} are valid just because of the Nahm equations
\begin{equation}\label{Nahm-class-s}
[p,v_a]_{{}_D}={\textstyle\frac12}\,\epsilon_{abc}\,[v_b,v_c]_{{}_D}
\end{equation}
for the bosonic variables $v_a$.
Indeed, the generators (\ref{superQ-v}) have the following Dirac brackets
\begin{eqnarray}\nonumber
\{Q^{i},\bar{Q}^{j}\}_{{}_D} &\!=\!& 2i\,\epsilon^{ij}\Big[  {\textstyle\frac{1}{4}} \,p^{2} +
{\textstyle\frac{1}{4}}
\left(v^{a}+c^{a}\right)\left(v^{\phantom{i}}_{a}+c_{a}\right)+
{\textstyle\frac12}\Big( [p,v_a]_{{}_D}+{\textstyle\frac12}\,
\epsilon_{abc}\,[v_b,v_c]_{{}_D}\Big)\chi_k \sigma_a^{kl}\bar\chi_l\Big]\\[4pt]
&\!\!& +\,i\,\sigma_a^{ij}\,\Big( [p,v_a]_{{}_D}-{\textstyle\frac12}\,
\epsilon_{abc}\,[v_b,v_c]_{{}_D}\Big)\,\chi^n \bar\chi_n\,,\label{QQ-cl-full}\\[4pt]
\{Q^{i},Q^{j}\}_{{}_D} &\!=\!& i\,\sigma_a^{ij}\,\Big( [p,v_a]_{{}_D}-{\textstyle\frac12}\,
\epsilon_{abc}\,[v_b,v_c]_{{}_D}\Big)\,\chi^n \chi_n\,.\nonumber
\end{eqnarray}
{}From these expressions we observe that the ${\cal N}=4$ supersymmetry algebra \p{susy-alg} takes place only
provided the equations (\ref{Nahm-class-s}) are valid.
As noticed in section\,3.2, in the considered system the validity of the equations (\ref{Nahm-class-s})
is a consequence of the Dirac brackets (\ref{DBxp-lin})-(\ref{DBvv-lin}).

Quantum counterparts of the supercharges  (\ref{superQ-v}) are uniquely found to be
\begin{equation}\label{superQ-v-quant}
\hat Q^{i} = \hat p\, \hat \chi^{i} +  i\left(\hat v_{a} + c_{a}\right)\sigma_a^{ik}  \hat\chi_{k} \,,
\qquad
\hat{\bar{Q}}_{i} = \hat p\, \hat{\bar\chi}_{i} -  i\left(\hat v_{a} + c_{a}^{\phantom{i}}\right)\sigma_{a\,ik} \hat {\bar{\chi}}^{k}
\,,
\end{equation}
where
\begin{equation}\label{alg-xi-quant}
\{\hat\chi^{i},\hat{\bar{\chi}}_{k}\} = {\textstyle\frac{1}{2}}\hbar\,\delta^{\,i}_k\,.
\end{equation}
The anticommutators of the supercharges  (\ref{superQ-v-quant}) are as follows
\begin{eqnarray}\nonumber
\{\hat Q^{i},\hat{\bar{Q}}^{j}\}  &\!=\!& -2\hbar\,\epsilon^{ij}\Big[  {\textstyle\frac{1}{4}} \,\hat p^{2} +
{\textstyle\frac{1}{4}}
\left(\hat v^{a}+c^{a}\right)\left(\hat v^{\phantom{i}}_{a}+c_{a}\right)-
{\textstyle\frac{i}{2}}\hbar^{-1}\Big( [\hat p,\hat v_a]+{\textstyle\frac12}\,\epsilon_{abc}\,[\hat v_b,\hat v_c]\Big)\hat \chi_k \sigma_a^{kl}\hat{\bar\chi}_l\Big]\\[4pt]
&\!\!& +\,i\,\sigma_a^{ij}\,\Big( [\hat p,\hat v_a]-{\textstyle\frac12}\,\epsilon_{abc}\,[\hat v_b,\hat v_c]\Big)\,
\Big(\hat \chi^n \hat{\bar\chi}_n-{\textstyle\frac12}\,\hbar \Big)\,,\label{QQ-qu-full}\\[4pt]
\{\hat Q^{i},\hat Q^{j}\}  &\!=\!& i\,\sigma_a^{ij}\,\Big( [\hat p,\hat v_a] -
{\textstyle\frac12}\,\epsilon_{abc}\,[\hat v_b,\hat v_c] \Big)\,\hat \chi^n \hat \chi_n\,.\nonumber
\end{eqnarray}
Then the fulfillment of the basic supersymmetry relations at the quantum level,
\begin{equation}\label{susy-alg-quant}
\{\hat Q^{i},\hat{\bar{Q}}_{k}\} = 2 \hbar\,\delta^{\,i}_k \hat H\,,\qquad \{\hat Q^{i},\hat Q^{k}\} =0\,,
\end{equation}
requires the validity of the operator Nahm equations
\begin{equation}\label{op-equat-quant}
[\hat p,\hat v_{a}] ={\textstyle\frac12}\,\epsilon_{abc}[\hat v_{b},\hat v_{c}]\,.
\end{equation}
The relevant quantum Hamiltonian is uniquely determined to have the form
\begin{equation}\label{H-su-v-quant}
\hat H = {\textstyle\frac{1}{4}} \,\hat p^{2} + {\textstyle\frac{1}{4}}
\left(\hat v_{a}+c_{a}\right)\left(\hat v^{\phantom{i}}_{a}+c_{a}\right)
- i\hbar^{-1}\, [\hat p,\hat v_{a}]\hat\chi_i \sigma_a^{ik}\hat{\bar\chi}_k\,.
\end{equation}

Thus, quite similarly to the classical case, where the vector variables $v_a$ are obliged to satisfy
the classical Nahm equations \p{Nahm-class-s}, after quantization
the quantum operators  $\hat v_a$ must be subjected to the operator Nahm equations \p{op-equat-quant}.

It is the appropriate place here to make a short account of what we have observed.

First, like in the classical case, the operator Nahm equations \p{op-equat-quant}
guarantee the existence of the ${\cal N}{=}\,4$ supersymmetry at the quantum level
for the supercharges of the form \p{superQ-v-quant}.

Second, while quantizing such systems, one should require the preservation of the Nahm equations
(i.e., the passing from the equations \p{Nahm-class-s} with Dirac brackets to the operator equations \p{op-equat-quant}),
in addition to the standard procedure of replacing the Dirac brackets of the phase variables by (anti)commutators.

The second point requires some additional comments concerning the chosen scheme of passing
to the quantum theory.

The equations (\ref{op-equat-quant})
involve the operators $\hat p$ and $\hat v_{a}$ which obey a complicated commutation algebra
induced by the Dirac brackets (\ref{DBxp-lin})-(\ref{DBvv-lin}).
As was already explained in section\,3 for the bosonic limit, in order to simplify things
we should split the basic quantum variables into the two decoupled sectors:
the one formed by $\hat x$, $\hat p$ and the second (spinning) sector spanned by the operators
$\hat\ell_a$ (or $\hat\varphi$, $\hat\ell_3$) with transparent algebraic and geometric properties.
Then the operators $\hat v_a{=}\hat v_a(\hat x,\hat\ell_a)$ (or $\hat v_a{=}\hat v_a(\hat x,\hat\varphi, \hat\ell_3)$)
are composite quantities and they are constructed from the corresponding classical expressions
$v_a{=}v_a(x,\ell_a)$ (or $v_a{=}v_a(x,\varphi, \ell_3)$) by the appropriate ordering of the operators $\hat\ell_a$.
In this case, the equations (\ref{op-equat-quant}) take the form of the operator Nahm equations
\begin{equation}\label{op-equat-quant1}
\hbar\,\frac{\partial}{\partial\hat x}\,\hat v_{a} ={\textstyle\frac{i}{2}}\,\epsilon_{abc}[\hat v_{b},\hat v_{c}]\,.
\end{equation}
Since $\hat x$ commutes with all operators $\hat\ell_a$ (and with $\hat\varphi$, $\hat\ell_3$),
the left-hand side of the equations (\ref{op-equat-quant1}) is directly specified by the classical expressions for $v_a\,$,
up to the ordering of $\hat\ell_a$. Then one is forced to assume that the right-hand side of eqs. (\ref{op-equat-quant1})
is also uniquely determined by the Dirac brackets, i.e. $[\hat v_{a},\hat v_{b}]=i\hbar\,\widehat{[ v_{a}, v_{b}]_{{}_D}}$
for  $v_a{=}v_a(x,\ell_a)$.
But it is obvious that these severe conditions cannot be generically satisfied, as soon as we proceed
from the standard quantum relations between the quantities $\ell_a$,
$[\hat\ell_{a},\hat\ell_{b}]=i\hbar\,\widehat{[\ell_{a},\ell_{b}]_{{}_D}}\,$, as the basic ones.
Only in some special cases we can expect an agreement of the quantization of the composite vector $v_{a}$ with
the quantization of the ``elementary'' constituents $\ell_{a}$, when simultaneously passing from the classical Nahm equations
to their quantized version. This becomes possible in the one-monopole case and special multi-monopole case, when
the components of the vector $v_{a}$ are linear functions of $\ell_{a}$.

In the standard two-monopole case, when the expansion of $v_{a}$ contains all degrees of $\ell_{a}$,
the only possibility to preserve Nahm equations (and, hence, ${\cal N}{=}4$ superalgebra) at the quantum level is
to properly modify the vector operators $\hat v_a$ as compared with their classical expressions,
$$
\hat v_a(\hat x,\hat\ell_a)\quad\rightarrow\quad\hat v_a(\hat x,\hat\ell_a;\hbar)\,,
$$
and to require that the new operators satisfy the operator Nahm equations to all orders in $\hbar$.
In this case, the zero-order term in the $\hbar$ expansion of operators $\hat v_a(\hat x,\hat\ell_a;\hbar)$
is uniquely determined by the corresponding classical expressions\footnote{More precisely, the classical expressions
are the Weyl symbols of the zero-order terms.}, while the higher-order terms are iteratively found from the requirement that
the quantum Nahm equations hold.

It should be emphasized that in our consideration we strictly follow the standard ideology of passing
from the classical system to the corresponding quantum theory. If one could directly construct a quantum theory,
without any reference to the classical system, no the problem of compliance with these additional restrictions would
arise. However, we do not know how to proceed in the second way.

\subsection{The one-monopole case}

Let us firstly consider the case of one--monopole potential (\ref{pot-lin-1}).
We use the variables (\ref{ell-radial}) subjected to the constraint (\ref{ell-radial-sq})
and parametrizing the two-sphere.
The Dirac brackets of bosonic variables are as in (\ref{DBxp-lin1a})-(\ref{DBvv-lin1a}).
With the choice $\vec{c}=-\vec{k}$ (see (\ref{c-k})) the generators (\ref{superQ-v}), (\ref{H-su-v}) take the form
\begin{equation}\label{superQ1}
Q^{i} = p\, \chi^{i} +  i\, \frac{\ell_a\,\sigma_a^{ik}\chi_k}{x}   \,,
\qquad
\bar{Q}_{i} = p\, \bar{\chi}_{i} -  i\, \frac{\ell_a\, \sigma_{a\,ik} \bar\chi^k}{x}
\,,
\end{equation}
\begin{equation}\label{H-su1}
H = \frac{1}{4}\left(p^2 + \frac{\ell_a\ell_a}{x^2}\right) + \frac{\ell_a\,\chi_i \sigma_a^{ik}\bar\chi_k}{x^2}\,.
\end{equation}

As opposed to the pure bosonic case considered in section\,3, the vector $\ell_a$ has non-vanishing
Dirac brackets with the Hamiltonian (\ref{H-su1}) and, also, with the supercharges (\ref{superQ1}).
The transformations generated by ${\cal N}{=}4$ supercharges and the Hamiltonian are realized on $\ell_a$ as
\begin{equation}\label{var-f1}
\delta{\ell}_a= [\beta H+\varepsilon_i Q^{i}-\bar\varepsilon^i\bar{Q}_{i}, {\ell}_a]_{{}_D}
= \epsilon_{abc\,} \omega_{b\,} \ell_c\,.
\end{equation}
Here, $\omega_b= \beta\,\sigma_b^{ik}\chi_i \bar\chi_k/x^2-i\sigma_b^{ik}(\varepsilon_i\chi_k
- \bar\varepsilon_i\bar\chi_k)/x$ and $\beta, \varepsilon_i$ and $\bar\varepsilon^i$ are parameters of the $t$-translations and
supertranslations.  Thus the supersymmetry transformations of $\ell_a$, like its $H$-transformation,
are represented as SU(2) rotations with the field-dependent parameters. As a result,
the supersymmetry--transformed vector $\tilde{\ell}_a = \ell_a + \delta \ell_a + \ldots$ satisfies the same basic relations
(\ref{ell-radial-sq}) and (\ref{DBvv-lin1a}), i.e.
$\tilde{\ell}_a\tilde{\ell}_a=g^2$ and $[\tilde{\ell}_a,\tilde{\ell}_b]_{{}_D}= \epsilon_{abc}\tilde{\ell}_c$.

The fuzzy-sphere coordinate $\ell_a$ is a part of the triplet of the generators
\begin{equation}\label{J-su1}
J_a = \ell_a \, - \,\chi_i \sigma_a^{ik}\bar\chi_k\,,\qquad
[J_a,J_b]_{{}_D}= \epsilon_{abc\,}J_c\,,
\end{equation}
which commute with the Hamiltonian, $[H,J_a]_{{}_D}= 0$, and generate ${\rm SU}(2)$ transformations
acting on all doublet indices $i, k$:
\begin{equation}\label{DB-JHQ1}
[Q^{i},J_a]_{{}_D}= -{\textstyle\frac{i}{2}} \,\sigma_a^{ik\,} {Q}_{k}\,,\qquad
[\bar{Q}_{i},J_a]_{{}_D}= {\textstyle\frac{i}{2}} \,\sigma_{a\,ik\,} \bar{Q}^{k}\,.
\end{equation}
The generators  (\ref{J-su1}), together with the generators (\ref{superQ1}),  (\ref{H-su1})
and
\begin{equation}\label{I-su1}
\begin{array}{rcl}
 I_{1^\prime} &=& {\textstyle\frac{1}{2}}\,(\chi_k \chi^k+\bar\chi^k \bar\chi_k)\,,\\
I_{2^\prime} &=& -{\textstyle\frac{i}{2}}\,(\chi_k \chi^k-\bar\chi^k \bar\chi_k)\,,\\
I_{3^\prime} &=& -\chi_k \bar\chi^k \,,
\end{array}\qquad
[I_{a^\prime},I_{b^\prime}]_{{}_D}= \epsilon_{a^\prime b^\prime c^\prime}I_{c^\prime}\,.
\end{equation}
\begin{equation}\label{superS1}
S^{i} = -2\,x\, \chi^{i} + t\, Q^{i}  \,,
\qquad
\bar{S}_{i} = -2\,x\, \bar{\chi}_{i} + t\, \bar{Q}_{i}
\,,
\end{equation}
\begin{equation}\label{KD-su1}
K = x^2 -t\,xp +t^2 H\,,\qquad
D=-{\textstyle\frac{1}{2}}\,xp +t\,H\,,
\end{equation}
constitute the algebra of the ${\cal N}=4$ conformal supergroup ${\rm OSp}(4|2)\,$.  This supergroup
provides the full symmetry of the component action in the one-monopole case.
Note that, although the quantities  (\ref{J-su1}) form the algebra ${\rm SU}(2)$ and commute with the Hamiltonian,
they cannot be treated as the coordinates of the fuzzy sphere: no condition $J_a J_a=const$ is valid
and, moreover, such a condition would be not invariant under supersymmetry transformations,
in contrast to the invariant condition ${\ell}_a{\ell}_a=g^2$. Therefore,
in the full supersymmetric setting,  the bosonic fuzzy sphere is still spanned
by the same variables $\ell_a$.\footnote{Note that here we consider the model with the world-line supersymmetry.
In the case of supersymmetrization of the target space there arise target fuzzy supermanifolds,
e.g. the so-called fuzzy supersphere (see \cite{BCIMT,Has} and refs. therein).}

The preservation of the basic relations (\ref{ell-radial-sq}) and (\ref{DBvv-lin1a}) which define the fuzzy sphere
suggests the use of the standard $(2n+1)\times (2n+1)$ matrix realization for $\hat\ell_a\,$, where
$n \in \mathbb{Z}, \mathbb{Z} + {\textstyle\frac{1}{2}}$ is the spin of  SU(2) irrep (``fuzziness'').
It should be emphasized that in this case the quantum Nahm equations (\ref{op-equat-quant1}),
which are necessary for the implementation of ${\cal N}{=}\,4$ supersymmetry at the quantum level, are valid
because $\hat v_a{-}k_a=\hat \ell_a/\hat x$ for the fuzzy sphere in the one--monopole case.

Using the holomorphic representation for the fermionic operators
\begin{equation}\label{real-fermi}
\hat\chi^{i}=\chi^{i}\,,\qquad \hat{\bar{\chi}}_{k}= {\textstyle\frac{1}{2}}\hbar\,\partial/\partial\chi^{k}
\end{equation}
and taking into account that the quantum supercharges and Hamiltonian are $(2n+1)\times (2n+1)$ matrices,
we find that the wave function should have the form
\begin{equation}\label{swf}
\Psi^A(x,\chi^{i})=\phi^A(x)+\chi^{i\,}\psi^A_i(x)+\chi^{i}\chi_{i\,}\varphi^A(x)\,,
\end{equation}
where the external index $A=1, \ldots 2n$ is the index of the irreducible SU(2) representation
with the matrices $\hat\ell_a$ as generators.
It is easy to see that, with respect to the full ${\rm SU}(2)$ transformations generated by (\ref{J-su1}),
the bosonic wave functions $\phi^A(x)$ and $\varphi^A(x)$ form two spin $n$ SU(2) irreps,
while the fermionic wave functions $\psi^A_i(x)$ carry two SU(2) irreps, with SU(2) spins $n \pm {\textstyle\frac{1}{2}}$.

This result is in full agreement with the result obtained in \cite{FIL2}, where the spin variables were represented
by the gauged ({\bf 4,4,0}) multiplet. In contradistinction to the formulation in \cite{FIL2}, where the fuzzy sphere
coordinates are constructed as bilinear products of the ${\rm SU}(2)$ doublet component fields obeying
the oscillator algebra, i.e. as some secondary composite objects, the ({\bf 3,4,1}) spin multiplet
provides the description of the fuzzy sphere directly in terms of the non-abelian vector coordinates $\ell_a\,$, which
are treated now as ``elementary'' constituents.

\subsection{Multi-monopole cases}

We first consider the quantization of the standard two-monopole case with the potential of the form (\ref{pot-lin-2a}).

Inverting the relations (\ref{pot-lin-2a}), (\ref{ell3-def}), (\ref{phi-def}), we find the following expressions
for the components of the three-vector $v_a$ in terms of the dynamical variable $x$
and the spin ones $\varphi, \ell_3$:
\begin{equation}\label{vec-2m}
v_1=V(x,\ell_3)\,\cos\varphi\,, \qquad v_2=V(x,\ell_3)\,\sin\varphi\,, \qquad v_3=W(x,\ell_3)\,.
\end{equation}
The explicit form of the functions $V(x,\ell_3)$, $W(x,\ell_3)$ in two limiting cases considered in section 3.4
can be easily obtained from the expressions (\ref{v3-lin-2a-1}), (\ref{v12-lin-2a-1}) and (\ref{v3-lin-2a-2}),
(\ref{v12-lin-2a-2}).
But the explicit form of these functions still does not prompt us how to quantize the system.

As argued in \cite{Sm}, the basic step in passing to the quantum supercharges from the classical ones
is to perform the Weyl ordering of the latter. Following this prescription and employing the simple algebra
of the basic operators $\hat x$, $\hat p$, $\hat\varphi$, $\hat\ell_3$, we can make use of the Moyal bracket \cite{Mo,BerMar,Sm}
in our analysis.

The quantum expressions corresponding to the expressions (\ref{vec-2m}) are
\begin{equation}\label{vec-2m-op}
\hat v_\pm:=\hat v_1\pm i\hat v_2=\langle V(\hat x,\hat\ell_3)\,e^{\pm i\hat\varphi}\rangle_{{}_W}\,, \qquad \hat v_3=W(\hat x,\hat\ell_3)\,,
\end{equation}
where the brackets $\langle...\rangle_{{}_W}$ denote the Weyl ordering of the noncommutative operators $\hat\varphi$, $\hat\ell_3$.
At the same time, classical expressions (\ref{vec-2m}) should be Weyl symbols of the corresponding quantum quantities, i.e
\begin{equation}\label{vec-2F-tr}
\hat v_\pm(\hat\ell_3,\hat\varphi)={\textstyle \frac{1}{4\pi^2}}\int
d\alpha\, d\beta\,d\ell_3\, d\varphi\, V(x,\ell_3)\,e^{\pm i\varphi}\,e^{- i(\alpha\ell_3+\beta\varphi)}\,e^{i(\alpha\hat\ell_3+\beta\hat\varphi)}\,.
\end{equation}
The Weyl ordering of the products of various operators is accomplished by means of the Moyal bracket.
In our case, the Moyal brackets of any operators $\hat M$, $\hat N$ is defined by the following general formula
\begin{eqnarray}\label{Moyal-br}
\left[[\hat M, \hat N\}\right]^W &=& 2\sinh\left\{
\frac{\hbar}{4}\left( \frac{\partial^2}{\partial\chi^{(2)k}\partial\bar\chi^{(1)}_k}
-\frac{\partial^2}{\partial\chi^{(1)k}\partial\bar\chi^{(2)}_k}\right)\right.
\\
&&\left. + \frac{i\hbar}{2}\left( \frac{\partial^2}{\partial x^{(1)}\partial p^{(2)}} -\frac{\partial^2}{\partial x^{(2)}\partial p^{(1)}}\right)
+ \frac{i\hbar}{2}\left( \frac{\partial^2}{\partial \varphi^{(1)}\partial \ell_3^{(2)}}
-\frac{\partial^2}{\partial \varphi^{(2)}\partial \ell_3^{(1)}}\right) \right\}
\nonumber
\\
&&\left.\cdot\, M(x^{(1)},p^{(1)},\varphi^{(1)},\ell_3^{(1)},\chi^{(1)k},\bar\chi^{(1)}_k)
\,N(x^{(2)},p^{(2)},\varphi^{(2)},\ell_3^{(2)},\chi^{(2)k},\bar\chi^{(2)}_k) \right|_{(1)=(2)}
\nonumber
\\
&\equiv&i\hbar\, [M,N\}_{{}_M}\,.\nonumber
\end{eqnarray}
Note that the definition (\ref{vec-2F-tr}) implies the exact operator relations
\begin{equation}\label{vec-2m-op-W}
\hat v_\pm= e^{\pm i\hat\varphi/2}\,V(\hat x,\hat\ell_3)\,e^{\pm i\hat\varphi/2}
\,.
\end{equation}
They are obtained by rewriting the operator exponential in the integral (\ref{vec-2F-tr}) as
$e^{i(\alpha\hat\ell_3+\beta\hat\varphi)}=e^{i\alpha\beta\hbar/2}e^{i\beta\hat\varphi}e^{i\alpha\hat\ell_3}$
and then performing the appropriate  Fourier transforms. We have also used the relation
$F(\hat\ell_3)\,e^{i\alpha\hat\varphi}=e^{ i\alpha\hat\varphi}F(\hat\ell_3+\alpha \hbar)$.

As we already know, the implementation of ${\cal N}=4$ supersymmetry  at the quantum level in the present model
requires the operator Nahm equations (\ref{op-equat-quant}). Their fulfillment for the Weyl
ordered quantities is expressed as the Moyal-Nahm equations \footnote{Moyal deformations of Nahm equations were earlier considered
in \cite{GCPl} and \cite{BaF}.}
\begin{equation}\label{W-op-equat-quant}
[ p, v_{a}]_{{}_M} ={\textstyle\frac12}\,\epsilon_{abc}[ v_{b}, v_{c}]_{{}_M}\,.
\end{equation}
A direct calculation with using the Moyal bracket (\ref{Moyal-br}) shows that the Moyal brackets in the left-hand and
right-hand sides of the equation  (\ref{W-op-equat-quant}) do not match each other: the left-hand side
of (\ref{W-op-equat-quant}) coincides with the Dirac brackets,
\begin{equation}\label{W-op-equat-quant-a}
[ p, v_{a}]_{{}_M} =-\frac{\partial v_{a}}{\partial x}=[ p, v_{a}]_{{}_D} \,,
\end{equation}
while the right-hand side contains extra terms of the order $\hbar^2$ and higher. For example,
\begin{equation}\label{W-op-equat-quant-b}
i\hbar\,[ v_{3}, v_{\pm}]_{{}_M}=
-2\,v_{3}\sinh\left\{\frac{i\hbar}{2}\,
\frac{\overleftarrow{\partial} }{\partial \ell_3}\,\frac{\overrightarrow{\partial} }{\partial \varphi}\right\}v_{\pm}=
i\hbar\,[ v_{3}, v_{\pm}]_{{}_D}+{\cal O}(\hbar^3)\,.
\end{equation}
Thus we are led to modify the passing to the quantum theory.

Note that in the one-monopole case, when the function $W$ is linear in $\ell_3$, such a problem does not arise.
We have $W=(\ell_3/x)+k$, $V=\sqrt{g^2-\ell_3^2}/x$ in this case and
the Weyl-ordered operators (\ref{vec-2m-op}) (see (\ref{vec-2m-op-W})) are
\begin{equation}\nonumber
\hat v_3=(\hat \ell_3/\hat x)+k\,,\qquad \hat v_\pm=
e^{\pm i\hat\varphi/2}\,\sqrt{g^2-\hat\ell_3^2}\,e^{\pm i\hat\varphi/2}/\hat x\,.
\end{equation}
They obey SU(2) algebra
\begin{equation}\nonumber
[\hat v_+,\hat v_-]=2\hbar\,\hat v_3/\hat x\,,\qquad
[\hat v_3,\hat v_\pm]=\pm\hbar\,\hat v_\pm/\hat x\,.
\end{equation}
As a result, the operator Nahm equations (\ref{op-equat-quant1}) are nicely satisfied in this case.

In the two-monopole case we must modify the classical-quantum correspondence (\ref{vec-2F-tr}) to save
the most important equation (\ref{W-op-equat-quant}) which guarantees the ${\cal N}=4$ supersymmetry at the quantum level.
For this purpose we need to change the symbols of the quantum
operators $\hat v_a$. Instead of the symbols (\ref{vec-2m}), basically coinciding with the classical expressions,
we consider the following ones
\begin{equation}\label{vec-2m-mod}
v_\pm=\tilde V(x,\ell_3,\hbar)\,e^{\pm i\varphi}\,, \qquad v_3=\tilde W(x,\ell_3,\hbar)\,.
\end{equation}
The correspondence principle with the initial system is ensured by the coincidence of the first terms
in the expansion of  (\ref{vec-2m-mod}) in $\hbar$ with the expressions (\ref{vec-2m}):
\begin{equation}\label{exp-2m-mod}
\begin{array}{rcl}
\tilde V(x,\ell_3,\hbar)&=&V(x,\ell_3)+\hbar\, V_1(x,\ell_3)+\hbar^2\, V_2(x,\ell_3)+\cdots\,, \\[6pt]
\tilde W(x,\ell_3,\hbar)&=&W(x,\ell_3)+\hbar\, W_1(x,\ell_3)+\hbar^2\, W_2(x,\ell_3)+\cdots\,.
\end{array}
\end{equation}
The relevant  operators read
\begin{equation}\label{vec-2m-op-mod}
\hat v_\pm= e^{\pm i\hat\varphi/2}\,\tilde V(\hat x,\hat\ell_3,\hbar)\,e^{\pm i\hat\varphi/2}\,,\qquad
\hat v_3=\tilde W(\hat x,\hat\ell_3,\hbar) \,.
\end{equation}
Thus we propose to correct the quantum operators in higher orders in the expansion
in $\hbar$, in such a way that the full operator Nahm equations are satisfied, while
the limit $\hbar\to 0$ still yields the classical system.

The Moyal-Nahm equations (\ref{W-op-equat-quant}) for the symbols (\ref{vec-2m-mod}) or, what is the same,
the operator Nahm equations (\ref{op-equat-quant}) for the operators (\ref{vec-2m-op-mod}) now amount to
the equations for the coefficient functions $V_n(x,\ell_3)$ and $W_n(x,\ell_3)$. Solving these equation, we can
find the complete solutions for the quantum operators. In Appendix we present the general scheme of
finding the solutions as series in $\hbar$ and explicitly give first non-trivial orders.

Thus, we succeeded in constructing the quantum theory in the two-monopole case,
using the expressions (\ref{vec-2m-mod}) as Weyl symbols  of the quantum operators (\ref{vec-2m-op-mod}).
In this way, the fulfillment of the operator Nahm equations guarantees the ${\cal N}{=}4, d\,{=}1$ Poincar\'e supersymmetry.
The generators of the Poincar\'e supersymmetry are given by the expression (\ref{superQ-v-quant}) and (\ref{H-su-v-quant}),
in which $\hat v_\pm$ and $\hat v_3$ are given  by eqs. (\ref{vec-2m-op-mod}), with the functions $\tilde V$ and $\tilde W$
defined as series in $\hbar$. For example, using the expression (\ref{A-i}) obtained in Appendix we find the first terms
of the quantum supercharges (up to the first order in $E$ and up to the fourth order in $\hbar$) in the case {\bf i)} of
section 3.4:
\begin{equation}\label{Q-2-qu}
\begin{array}{lcl}
\hat Q^{i} &=& {\displaystyle\hat p\, \hat \chi^{i} }+{\displaystyle   i\left[(\tilde W -k )\sigma_3^{ik}
+  e^{i\hat\varphi/2}\tilde V e^{i\hat\varphi/2}\sigma_-^{ik}+  e^{-i\hat\varphi/2}\tilde V e^{-i\hat\varphi/2}\sigma_+^{ik}\right]\hat\chi_{k}}\\[9pt]
\hat{\bar{Q}}_{i} &=& {\displaystyle\hat p\, \hat{\bar\chi}_{i} }-{\displaystyle   i\left[(\tilde W -k )\sigma_{3\,ik}
+  e^{i\hat\varphi/2}\tilde V e^{i\hat\varphi/2}\sigma_{-\,ik}+  e^{-i\hat\varphi/2}\tilde V e^{-i\hat\varphi/2}\sigma_{+\,ik}\right]\hat {\bar{\chi}}^{k}},
\end{array}
\end{equation}
where $\sigma_{\pm}=\frac12\,(\sigma_{1}\pm i\sigma_{2})$ and
\begin{equation}\label{}
\begin{array}{rcl}
\tilde W (\hat x,\hat\ell_3)-k &=&{\displaystyle\frac{1}{\hat x}\left(\displaystyle \hat\ell_3 +
E\,\frac{3\hat\ell_3^2-g^2}{\hat x^2} \right)},\\[9pt]
\tilde V (\hat x,\hat\ell_3,\hbar) &=&{\displaystyle \frac{\sqrt{g^2-\hat\ell_3^2}}{\hat x}\left(1 +\frac{3E\hat\ell_3}{\hat x^2}\right)\left(1+\frac{\hbar^2}{8(g^2-\hat\ell_3^2)} +
\frac{\hbar^4}{128(g^2-\hat \ell_3^2)^{2}}\right) }.
\end{array}
\end{equation}
In (\ref{Q-2-qu}) we take $\vec{c}=-\vec{k}$ as in the one-center case.

There is also another way of constructing a quantum ${\cal N}{=}4$ supersymmetric system in the
multi-monopole case, which bears close parallels with the fuzzy-sphere method of the one-monopole case.
This option is associated with the special multi-monopole system considered in section 3.3.
We just consider the quantum counterparts of the relations (\ref{new-prop-1})
\begin{equation}\label{new-prop-1q}
\hat v_1=f_1(\hat x)\,\hat \ell_{1}\,, \qquad \hat v_2=f_2(\hat x)\,\hat \ell_{2}\,, \qquad \hat v_3=f_3(\hat x)\,\hat \ell_{3}\,,
\end{equation}
where $\hat \ell_{a}$ are the standard fuzzy sphere coordinates,
\begin{equation}\label{LL-lin3a}
[\hat\ell_a,\hat\ell_b]= i\hbar\,\epsilon_{abc}\hat\ell_c\,,\qquad
\hat\ell_{a}\hat\ell_{a} =g^2\,,
\end{equation}
and $f_a$ are defined in (\ref{sol-f123}).
As pointed out in section\,3.1., we must make the identification
$g^2 = \hbar^2\, n(n+1)$, $2n\,{\in}\,\mathbb{N}$ to deal with the unitary SU(2) representations.

Due to the Euler equations (\ref{new-eq-1}) quantum Nahm equations (\ref{op-equat-quant}), (\ref{op-equat-quant1}) are satisfied.
As a result, the operators (we choose $c_a=(0,0,c)$ in (\ref{superQ-v-quant}), (\ref{H-su-v-quant}))
\begin{equation}\label{superQ-v-quant-1}
\begin{array}{rcl}
\hat Q^{i} &=& \hat p\, \hat \chi^{i}
 +  \frac{ik}{g}\,{\rm{sinh}}^{-1}(\frac{k\hat x}{g})
\left[ \hat\ell_1 \, \sigma_1^{ik}\hat{\chi}_k +
\hat\ell_2\, \sigma_2^{ik}\hat{\chi}_k \right. \\[7pt]
&& \qquad \qquad\qquad \qquad\qquad + \left.
\left({\rm{cosh}}(\frac{k\hat x}{g})\,\hat\ell_3+\frac{cg}{k}\,{\rm{sinh}}(\frac{k\hat x}{g})\right) \sigma_3^{ik}\hat{\chi}_k\right]  \,, \\[7pt]
\hat{\bar{Q}}_{i} &=& \hat p\, \hat{\bar\chi}_{i} -
\frac{ik}{g}\,{\rm{sinh}}^{-1}(\frac{k\hat x}{g})
\left[ \hat\ell_1 \, \sigma_{1\,ik}\hat{\bar\chi}^k +
\hat\ell_2\, \sigma_{2\,ik}\hat{\bar\chi}^k \right. \\[7pt]
&& \qquad \qquad\qquad \qquad\qquad + \left.
\left({\rm{cosh}}(\frac{k\hat x}{g})\,\hat\ell_3+\frac{cg}{k}\,{\rm{sinh}}(\frac{k\hat x}{g})\right) \sigma_{3\,ik}\hat{\bar\chi}^k\right]
\,,
\end{array}
\end{equation}
\begin{equation}\label{H-su-v-quant-1}
\begin{array}{rcl}
\hat H &=& \frac{1}{4}\,\hat p^{2}  + \frac{k^2}{4g^2}\,{\rm{sinh}}^{-2}(\frac{k\hat x}{g})
\left[ (\hat\ell_1)^2+(\hat\ell_2)^2+
\left({\rm{cosh}}(\frac{k\hat x}{g})\,\hat\ell_3+\frac{cg}{k}\,{\rm{sinh}}(\frac{k\hat x}{g})\right)^2\right] \\[10pt]
&& + \frac{k^2}{g^2}\,{\rm{sinh}}^{-2}(\frac{k\hat x}{g})
\left[ {\rm{cosh}}^{2}(\frac{k\hat x}{g})\left(\hat\ell_1 \hat\chi_i \sigma_1^{ik}\hat{\bar\chi}_k +
\hat\ell_2\hat\chi_i \sigma_2^{ik}\hat{\bar\chi}_k\right)+\hat\ell_3\hat\chi_i \sigma_3^{ik}\hat{\bar\chi}_k\right]
\end{array}
\end{equation}
form the standard ${\cal N}{=}4, d\,{=}1$ Poincar\'{e} superalgebra.

\setcounter{equation}0
\section{Summary and outlook}

In this paper we presented a new version of ${\cal N}{=}\,4$ mechanics,
which couples a ({\bf 1,4,3}) multiplet to a ({\bf 3,4,1}) multiplet.
The ({\bf 1,4,3}) multiplet represents one dynamical bosonic and four dynamical fermionic variables.
The ({\bf 3,4,1}) multiplet appears in a superfield Wess--Zumino action and thus is ``semi-dynamical'';
after elimination of the auxiliary fermions a bosonic three-vector spin variable remains.
The ${\cal N}{=}\,4$ supersymmetric coupling of the multiplets generates a constraint
which relates one degree of freedom of these vector variables to the dynamical boson.
The remaining two bosons of the semi-dynamical ({\bf 3,4,1}) multiplet are genuine spin variables.

These spin variables parametrize some two-dimensional fuzzy surface in three-dimensional (flat) space.
The Dirac brackets defined by the harmonic scalar potential yield an algebraic structure in the spin sector.
For the one-center potential (\ref{pot-lin-1}) and the special multi-center potential (\ref{pot-tilde}),
the spin variables form an SU(2) algebra and parametrize the fuzzy two-sphere.
These quantum models can be given in closed form, while the one with a general two-center potential
needs a power series expansion (in small parameters and in~$\hbar\,$).

An unexpected and, in our opinion, most remarkable feature is the occurrence of the Nahm equations
for the three-vector spin variable as a consequence of the Dirac brackets of the constraints.
We discovered a strict correspondence between these Nahm equations and the presence of ${\cal N}{=}\,4$
supersymmetry in the model, classically and quantum mechanically.
In other words, the Nahm equations guarantee extended supersymmetry.

We did not yet study the most general type of models possible. Rather, we restricted ourselves to the
action~(\ref{ac-bcom-lin1}) and to special multi-monopole configurations.
It would be interesting to study the general multi-center solution of the Laplace equation
$\partial_a\partial_a\,\mathscr{U}=0$, which is given by~(\ref{pot-lin-general}).
For this case one may expect the spin variables to parametrize some fuzzy Riemann surface
(see, e.g.,~\cite{ABHHS,GPS}) and form a nonlinear deformed algebra (see~\cite{BHTjin} and references therein).
Furthermore, our supersymmetry generators are linear in the fermionic variables, which is also special.
In the more general case of ${\cal N}{=}\,4$ supersymmetry generators cubic in the fermions
the Nahm equations might get supplemented by additional relations to ensure full extended supersymmetry.
Finally, it remains to investigate other combinations of dynamical and semi-dynamical ${\cal N}{=}\,4$ multiplets
for describing spin variables, utilizing for instance the nonlinear ({\bf 3,4,1}) multiplet~\cite{IL,IKLecht2}.

\bigskip
\section*{Acknowledgements}

\noindent
The authors would like to thank M.\,Konyushikhin and A.\,Smilga for valuable remarks
and V.\,Kolontsov for collaboration at an early stage of this work.
We acknowledge support from a grant of the Heisenberg-Landau Programme,
RFBR grants 09-01-93107, 11-02-90445, 12-02-00517 (S.F.~\&~E.I.)
and a DFG grant, project no.~436 RUS/113/669 (E.I.~\&~O.L.).
S.F.~\&~E.I.\ would like to thank the Institute of Theoretical Physics at Leibniz
University of Hannover for its warm hospitality at different stages of this study.
E.I.\ thanks SUBATECH, Universit\'e de Nantes, for its kind hospitality at the final stage.

\bigskip

\renewcommand\theequation{A.\arabic{equation}} \setcounter{equation}0
\section*{Appendix \quad  }

In this Appendix we find the solution of the Nahm equations for the quantum-modified functions
\begin{equation}\label{A-vec-2m-op-mod}
\hat v_\pm= e^{\pm i\hat\varphi/2}\,\tilde V(\hat x,\hat\ell_3,\hbar)\,e^{\pm i\hat\varphi/2}\,,\qquad
\hat v_3=\tilde W(\hat x,\hat\ell_3,\hbar) \,,
\end{equation}
where
\begin{equation}\label{A-exp-2m-mod}
\begin{array}{rcl}
\tilde V(x,\ell_3,\hbar)&=&V_0(x,\ell_3)+\hbar^2\, V_2(x,\ell_3)+\hbar^4\, V_4(x,\ell_3)+\cdots\,, \\[6pt]
\tilde W(x,\ell_3,\hbar)&=&W_0(x,\ell_3)+\hbar^2\, W_2(x,\ell_3)+\hbar^4\, W_4(x,\ell_3)+\cdots\,.
\end{array}
\end{equation}

The operator Nahm equations
\begin{equation}\label{A-M-equat-quant}
[ \hat p, \hat v_{a}] ={\textstyle\frac12}\,\epsilon_{abc}[ \hat v_{b}, \hat v_{c}]\,,
\end{equation}
or their corresponding Moyal representation
\begin{equation}\label{A-MN-equat-quant}
[ p, v_{a}]_{{}_M} ={\textstyle\frac12}\,\epsilon_{abc}[ v_{b}, v_{c}]_{{}_M}\,,
\end{equation}
amount to the following equations for the functions (\ref{A-exp-2m-mod})
\begin{equation}\label{A-eqs-W}
\hbar\,\partial_x \tilde W(\ell_3)={\textstyle\frac{1}{2}}\left[ \tilde V^2(\ell_3 +\hbar/2)-\tilde V^2(\ell_3 -\hbar/2)\right]\,,
\end{equation}
\begin{equation}\label{A-eqs-V}
\hbar\,\partial_x \tilde V(\ell_3)=-\left[ \tilde W(\ell_3 +\hbar/2)-\tilde W(\ell_3 -\hbar/2)\right]V(\ell_3)\,.
\end{equation}

Since the Nahm equations with the Moyal bracket must be corrected at the level $\hbar^2$ and even higher-order levels,
in the expansions (\ref{A-exp-2m-mod}) we assume that $V_n(x,\ell_3)=W_n(x,\ell_3)=0\,$ for $n=2k+1$.
Then, the equations (\ref{A-eqs-W}), (\ref{A-eqs-V}) give rise to an infinite set of the equations
for the coefficient functions in the $\hbar^2$\,-expansion:
\begin{equation}\label{A-eqs-0}
\frac{\partial V_0}{\partial x}=-V_0\,\frac{\partial W_0}{\partial \ell_3}\,,\qquad
\frac{\partial W_0}{\partial x}=V_0\,\frac{\partial V_0}{\partial \ell_3}\,,
\end{equation}
\begin{equation}\label{A-eqs-2}
\frac{\partial V_2}{\partial x}=-V_2\,\frac{\partial W_0}{\partial \ell_3}-V_0\,\frac{\partial W_2}{\partial \ell_3}
-\frac{V_0}{2^2 3!}\,\frac{\partial^3 W_0}{\partial \ell_3{}^3}\,,\qquad
\frac{\partial W_2}{\partial x}=\frac{\partial (V_0V_2)}{\partial \ell_3}+\frac{1}{2^3 3!}\,
\frac{\partial^3 (V_0{}^2)}{\partial \ell_3{}^3}\,,
\end{equation}
\begin{equation}\label{A-eqs-4}
\frac{\partial V_4}{\partial x}=-V_4\,\frac{\partial W_0}{\partial \ell_3}-V_2\left(\frac{\partial W_2}{\partial \ell_3}+
\frac{1}{2^2 3!}\frac{\partial^3 W_0}{\partial \ell_3{}^3}\right)
-V\left(\frac{\partial W_4}{\partial \ell_3} +
\frac{1}{2^2 3!}\,\frac{\partial^3 W_2}{\partial \ell_3{}^3}+
\frac{1}{2^4 5!}\,\frac{\partial^5 W_0}{\partial \ell_3{}^5}\right),\qquad\qquad\qquad
\end{equation}
\begin{equation}\nonumber
\qquad\qquad\qquad\qquad
\frac{\partial W_4}{\partial x}=V_2\frac{\partial V_2}{\partial \ell_3}+\frac{\partial (V_0V_4)}{\partial \ell_3}
+\frac{1}{2^2 3!}\,\frac{\partial^3 (V_0V_2)}{\partial \ell_3{}^3}+\frac{1}{2^5 5!}\,
\frac{\partial^5 (V_0{}^2)}{\partial \ell_3{}^5}\,,
\end{equation}
$$
\quad\cdots\quad\,, \qquad\quad\cdots\;.
$$

The equations \p{A-eqs-0} are automatically satisfied by the classical expressions \p{v3-lin-2a-1}, \p{v12-lin-2a-1}
and (\ref{v3-lin-2a-2}), (\ref{v12-lin-2a-2}) for the two limiting cases considered in section 3.3:
\begin{equation}
\mbox{\bf i)}\qquad\left\{
\begin{array}{rcl}
V_0&=&{\displaystyle\frac{\sqrt{g^2-\ell_3^2}}{x}\left(1 +E\,\frac{3\ell_3}{x^2}\right) +{\cal O}(E^2),}\\[9pt]
W_0&=&{\displaystyle k+\frac{\ell_3}{x} +E\,\frac{3\ell_3^2-g^2}{x^3} +{\cal O}(E^2);}
\end{array}
\right.
\end{equation}
\begin{equation}
\mbox{\bf ii)}\qquad\left\{
\begin{array}{rcl}
V_0&=&{\displaystyle\frac{\sqrt{g^2-\ell_3^2}}{x} \left(1+d^2\,\frac{x^2}{2g^4}-d^3\,\frac{2x^3\ell_3}{g^7}\right) +{\cal O}(d^4)\,,}\\[9pt]
W_0&=&{\displaystyle k+\frac{\ell_3}{x} +d\,\frac{1}{g} -d^2\,\frac{x\ell_3}{g^4}+d^3\,\frac{x^2(3\ell_3^2-g^2)}{g^7} +{\cal O}(d^4)\,.}
\end{array}
\right.
\end{equation}
In fact, these equations are just the classical Nahm equations \p{Nahm-class-s} with Dirac brackets.

The remaining differential equations  \p{A-eqs-2}, \p{A-eqs-4}, and so on, serve to define the functions $V_2,V_4,\ldots$
and $W_2,W_4,\ldots$, respectively.
Note that it is enough to take only some particular solution of these equations to obtain a
self-consistent quantum system corresponding to the given classical system.
In the cases considered here the function $W_0$ has the degree two in $\ell_3$, and it already induces the first
nontrivial quantum corrections in $V_n$ and $W_n$ with $n>0$.
The expressions  for the leading in $\hbar^2$ and $\hbar^4$ components are as follows
\begin{equation}
\!\mbox{\bf i)}\quad\left\{
\begin{array}{lcl}
V_2={\displaystyle\frac{1}{8x(g^2-\ell_3^2)^{1/2}}\left(1 +E\,\frac{3\ell_3}{x^2}\right) +{\cal O}(E^2),}\quad &&
W_2= {\cal O}(E^2),\\[9pt]
V_4={\displaystyle\frac{1}{128x(g^2-\ell_3^2)^{3/2}}\left(1 +E\,\frac{3\ell_3}{x^2}\right) +{\cal O}(E^2),}\quad &&
W_4= {\cal O}(E^2);
\end{array}
\right.
\end{equation}
\begin{equation}
\!\mbox{\bf ii)}\quad\left\{
\begin{array}{rcl}
V_2={\displaystyle\frac{1}{8x(g^2-\ell_3^2)^{1/2}} \left(1+d^2\,\frac{x^2}{2g^4}-d^3\,\frac{2x^3\ell_3}{g^7}\right) +{\cal O}(d^4),}\quad &&
W_2= {\cal O}(d^4),\\[9pt]
V_4={\displaystyle\frac{1}{128x(g^2-\ell_3^2)^{3/2}} \left(1+d^2\,\frac{x^2}{2g^4}-d^3\,\frac{2x^3\ell_3}{g^7}\right) +
{\cal O}(d^4),}\quad &&
W_4= {\cal O}(d^4).
\end{array}
\right.
\end{equation}
As a result, we find the solutions up to the $\hbar^4$ terms
\begin{equation}\label{A-i}
\!\!\!\mbox{\bf i)}\quad\left\{
\begin{array}{lcl}
\tilde V\!\!\!\!&=&\!\!\!{\displaystyle\frac{\sqrt{g^2-\ell_3^2}}{x}\left(1 +\frac{3E\ell_3}{x^2}\right)\left(1+\frac{\hbar^2}{8(g^2-\ell_3^2)} +
\frac{\hbar^4}{128(g^2-\ell_3^2)^{2}}\right)+\,{\cal O}(E^2,\hbar^5) }\\[9pt]
\tilde W\!\!\!\!&=&\!\!\!{ \displaystyle k+\frac{\ell_3}{x} +E\,\frac{3\ell_3^2-g^2}{x^3} +{\cal O}(E^2,\hbar^5)};
\end{array}
\right.
\end{equation}
\begin{equation}\label{A-ii}
\!\!\!\!\!\!\mbox{\bf ii)}\quad\left\{
\begin{array}{rcl}
\tilde V&=&{\displaystyle\frac{\sqrt{g^2-\ell_3^2}}{x} \left(1+d^2\,\frac{x^2}{2g^4}-d^3\,\frac{2x^3\ell_3}{g^7}\right)
\left(1+\frac{\hbar^2}{8(g^2-\ell_3^2)} +
\frac{\hbar^4}{128(g^2-\ell_3^2)^{2}}\right)}  \\[9pt]
&&\hspace{10,5cm}+\,{\cal O}(d^4,\hbar^5),\\[6pt]
\tilde W&=&{\displaystyle k+\frac{\ell_3}{x} +d\,\frac{1}{g} -d^2\,\frac{x\ell_3}{g^4}+d^3\,\frac{x^2(3\ell_3^2-g^2)}{g^7}
+{\cal O}(d^4,\hbar^5)}.
\end{array}
\right.
\end{equation}

%\newpage


\begin{thebibliography}{96}
\addtolength{\itemsep}{-5.6pt}

\bibitem{FIL1}
S.\,Fedoruk, E.\,Ivanov, O.\,Lechtenfeld, {\it Supersymmetric Calogero models by gauging}, \\
Phys. Rev. {\bf D79} (2009) 105015, {\tt arXiv:0812.4276 [hep-th]}.

\bibitem{FIL2}
S.\,Fedoruk, E.\,Ivanov, O.\,Lechtenfeld, {\it OSp(4$|$2) superconformal mechanics}, \\
JHEP {\bf 0908} (2009) 081, {\tt arXiv:0905.4951\,[hep-th]}.

\bibitem{FIL3}
S.\,Fedoruk, E.\,Ivanov, O.\,Lechtenfeld, {\it New D(2,1; $\alpha$) mechanics with spin variables}, \\
JHEP {\bf 1004} (2010) 129, {\tt arXiv:0912.3508 [hep-th]}.

\bibitem{FIL-r}
S.\,Fedoruk, E.\,Ivanov, O.\,Lechtenfeld, {\it Superconformal mechanics}, \\
J. Phys. A: Math. Theor. {\bf 45} (2012) 173001, {\tt arXiv:1112.1947 [hep-th]}.

\bibitem{Mad}
J.\,Madore, {\it Quantum mechanics on a fuzzy sphere},
Phys. Lett. {\bf B263} (1991) 245;\\
{\it The fuzzy sphere}, Class. Quant. Grav. {\bf 9} (1992) 69.

\bibitem{BK1}
S.\,Bellucci, S.\,Krivonos, {\it Potentials in N{=}4 superconformal mechanics},\\
Phys. Rev. {\bf D80} (2009) 065022, {\tt arXiv:0905.4633\,[hep-th]}.

\bibitem{KL}
S.\,Krivonos, O.\,Lechtenfeld,
{\it SU(2) reduction in N=4 supersymmetric mechanics}, \\
Phys. Rev. {\bf D80} (2009) 045019, {\tt arXiv:0906.2469 [hep-th]}.

\bibitem{KL3}
S.\,Krivonos, O.\,Lechtenfeld,\\
{\it Many-particle mechanics with $D(2,1;\alpha)$ superconformal symmetry,}\\
JHEP {\bf 1102} (2011) 042, {\tt arXiv:1012.4639\,[hep-th]}.

\bibitem{KL1}
T.\,Hakobyan, S.\,Krivonos, O.\,Lechtenfeld, A.\,Nersessian,\\
{\it Hidden symmetries of integrable conformal mechanical systems},\\
Phys. Lett. {\bf A374} (2010) 801, {\tt arXiv:0908.3290 [hep-th]}.

\bibitem{BKS} 
S.\,Bellucci, S.\,Krivonos, A.\,Sutulin
{\it Three dimensional N=4 supersymmetric mechanics with Wu-Yang monopole},\\
Phys. Rev. {\bf D81} (2010) 105026, {\tt arXiv:0911.3257 [hep-th]}.

\bibitem{ISKon}
E.A.\,Ivanov, M.A.\,Konyushikhin, A.V.\,Smilga, \\
{\it SQM with non-Abelian self-dual fields: harmonic superspace description},\\
JHEP {\bf 1005} (2010) 003, {\tt arXiv:0912.3289 [hep-th]}.

\bibitem{KL2} 
S.\,Krivonos, O.\,Lechtenfeld, A.\,Sutulin,
{\it N=4 Supersymmetry and the BPST Instanton},
Phys. Rev. {\bf D81} (2010) 085021, {\tt arXiv:1001.2659 [hep-th]}.

\bibitem{IKon}
E.\,Ivanov, M.\,Konyushikhin, \\
{\it  N=4, 3D supersymmetric quantum mechanics in non-Abelian monopole background},\\
Phys. Rev. {\bf D82} (2010) 085014, {\tt arXiv:1004.4597 [hep-th]}.

\bibitem{GIOS}
A.S.\,Galperin, E.A.\,Ivanov, V.I.\,Ogievetsky, E.S.\,Sokatchev, \\ {\it Harmonic Superspace},
Cambridge Univ. Press, 2001.

\bibitem{IL}
E.\,Ivanov, O.\,Lechtenfeld, {\it N=4 supersymmetric mechanics in harmonic superspace},\\
JHEP {\bf 0309} (2003) 073, {\tt arXiv:hep-th/0307111}.

\bibitem{DI1}
F.\,Delduc, E.\,Ivanov, \\
{\it Gauging N=4 supersymmetric mechanics II: (1,4,3) models from the (4,4,0) ones}, \\
Nucl. Phys. {\bf B770} (2007) 179, {\tt arXiv:hep-th/0611247}.

\bibitem{DI}
F.\,Delduc, E.\,Ivanov, {\it Gauging N=4 supersymmetric mechanics}, \\
Nucl. Phys. {\bf B753} (2006) 211, {\tt arXiv:hep-th/0605211}.

\bibitem{IKL}
E.A.\,Ivanov, S.O.\,Krivonos, V.M.\,Leviant, \\ {\it Geometric superfield approach to
superconformal mechanics}, J. Phys. {\bf A22} (1989) 4201.

\bibitem{FJ}
L.\,Faddeev, R.\,Jackiw,
{\it Hamiltonian reduction of unconstrained and constrained systems},
Phys. Rev. Lett. {\bf 60} (1988) 1692;\\
G.V.\,Dunne, R.\,Jackiw, C.A.\,Trugenberger,\\
{\it ``Topological'' (Chern-Simons) quantum mechanics},
Phys. Rev. {\bf D41} (1990) 661.

\bibitem{FRS}
R.\,Floreanini, R.\,Percacci, E.\,Sezgin,\\
{\it Sigma models with purely Wess-Zumino-Witten actions},
Nucl. Phys. {\bf B322} (1989) 255; \\
{\it Infinite dimensional algebras in Chern-Simons quantum mechanics},\\
Phys. Lett. {\bf B261} (1991) 51, {\tt arXiv:hep-th/9111053}.

\bibitem{HTown}
P.S.\,Howe,  P.K.\,Townsend,
{\it The massless superparticle as Chern-Simons mechanics}, \\
Phys. Lett. {\bf B259} (1991) 285.

\bibitem{EGHan}
T.\,Eguchi, P.B.\,Gilkey, A.J.\,Hanson, \\ {\it Gravitation, Gauge Theories and Differential Geometry},
Phys. Rept. {\bf 66} (1980) 213.

\bibitem{IKLecht} E.\,Ivanov, S.\,Krivonos, O.\,Lechtenfeld,
{\it New variant of N=4 superconformal mechanics},\\
JHEP {\bf 0303} (2003) 014, {\tt arXiv:hep-th/0212303}.

\bibitem{Nahm}
W.\,Nahm, {\it The algebraic geometry of multimonopoles},
Lect. Notes Phys. {\bf 180} (1983) 456;\\
N.J.\,Hitchin, {\it On the construction of monopoles},
Commun. Math. Phys. {\bf 89} (1983) 145;\\
S.K.\,Donaldson, {\it Instantons and geometric invariant theory},\\
Commun. Math. Phys. {\bf 93} (1984) 453.

\bibitem{Ward}
R.S.\,Ward, {\it Linearization of the $SU(\infty)$ Nahm equations},
Phys. Lett. {\bf B234} (1990) 81.

\bibitem{GCPl}
H.\,Garcia-Compean, J.F.\,Plebanski,\\
{\it On the Weyl-Wigner-Moyal description of $SU(\infty)$ Nahm equations},\\
Phys. Lett. {\bf A234} (1997) 5, {\tt arXiv:hep-th/9612221}.

\bibitem{Dun}
M.\,Dunajski, {\it Harmonic functions, central quadrics, and twistor theory},\\
Class. Quant. Grav. {\bf 20} (2003) 3427,
{\tt arXiv:math/0303181}.

\bibitem{Per}
A.\,Perelomov,
{\it Generalized coherent states and their applications},
Springer, Berlin, 1986.

\bibitem{IM}
E.\,Ivanov, L.\,Mezincescu, P.K.\,Townsend, {\it Fuzzy $CP(n|m)$ as a quantum superspace},\\
Contribution to ``Symmetries in Gravity and Field Theory",
conference for J.-A. de Azcarraga's 60th birthday, June 2003, Salamanca, Spain,
{\tt arXiv:hep-th/0311159};\\
L.\,Mezincescu, {\it Super Chern-Simons quantum mechanics},\\
Proceedings of the International Workshop ``Supersymmetries and Quantum Symmetries"
(SQS'03, 24-29 July, 2003), Dubna, Russia,
{\tt arXiv:hep-th/0405031}.

\bibitem{GORV}
G.W.\,Gibbons, D.\,Olivier, P.J.\,Ruback, G.\,Valent, \\ {\it Multicenter metrics and harmonic superspace},
Nucl. Phys. {\bf B296} (1988) 679.

\bibitem{Gib}
G.W.\,Gibbons, {\it Gravitational instantons, confocal quadrics and separability of the \\
Schr\"{o}dinger and Hamilton-Jacobi equations},\\
Class. Quant. Grav. {\bf 20} (2003) 4401, {\tt arXiv:math/0303191}.

\bibitem{BCIMT}
A.\,Beylin, T.L.\,Curtright, E.\,Ivanov, L.\,Mezincescu, P.K.\,Townsend,\\
{\it Unitary spherical super-Landau models}, \\
JHEP {\bf 0810} (2008) 069, {\tt arXiv:0806.4716 [hep-th]}.

\bibitem{Has}
K.\,Hasebe,  {\it Hopf maps, lowest Landau level, and fuzzy spheres},\\
SIGMA {\bf 6} (2010) 071, {\tt arXiv:1009.1192 [hep-th]}.

\bibitem{Sm}
A.V.\,Smilga,
{\it How to quantize supersymmetric theories},
Nucl. Phys. {\bf B292} (1987) 363.

\bibitem{Mo}
J.E.\,Moyal,
{\it Quantum mechanics as a statistical theory},\\
Proc. Cambridge Phil. Soc. {\bf 45} (1949) 99.

\bibitem{BerMar}
F.A.\,Berezin, M.S.\,Marinov,\\
{\it Particle spin dynamics as the Grassmann variant of classical mechanics},\\
Annals Phys. {\bf 104} (1977) 336.

\bibitem{BaF}
L.\,Baker, D.\,Fairlie, {\it Moyal Nahm equations},\\
J. Math. Phys. {\bf 40} (1999) 2539, {\tt arXiv:hep-th/9901072}.

\bibitem{ABHHS}
J.\,Arnlind, M.\,Bordemann, L.\,Hofer, J.\,Hoppe, H.\,Shimada,
{\it Fuzzy Riemann surfaces}, \\ JHEP {\bf 0906} (2009) 047, {\tt arXiv:hep-th/0602290}.

\bibitem{GPS}
T.R.\,Govindarajan, P.\,Padmanabhan, T.\,Shreecharan,
{\it Beyond fuzzy spheres}, \\
J. Phys. {\bf A43} (2010) 205203, {\tt arXiv:0906.1660 [hep-th]}.

\bibitem{BHTjin}
J.\,de\,Boer, F.\,Harmsze, T.\,Tjin,\\
{\it Non-linear finite $W$-symmetries and applications in elementary systems}, \\
Phys. Rept. {\bf 272} (1996) 139, {\tt arXiv:hep-th/9503161}.

\bibitem{IKLecht2}
E.\,Ivanov, S.\,Krivonos, O.\,Lechtenfeld,\\
{\it N=4, d=1 supermultiplets from nonlinear realizations of $D(2,1;\alpha)$},\\
Class. Quant. Grav. {\bf 21} (2004) 1031, {\tt arXiv:hep-th/0310299}.


\end{thebibliography}
\end{document}